# Physical Layer Authentication Based on Hierarchical Variational Auto-Encoder for Industrial Internet of Things

Rui Meng, Xiaodong Xu, *Senior Member, IEEE*, Bizhu Wang, Hao Sun, Shida Xia, Shujun Han, *Member, IEEE*, and Ping Zhang, *Fellow, IEEE*

***Abstract*—Recently, Physical Layer Authentication (PLA) has attracted much attention since it takes advantage of the channel randomness nature of transmission media to achieve communication confidentiality and authentication. In the complex environment, such as the Industrial Internet of Things (IIoT), machine learning (ML) is widely employed with PLA to extract and analyze complex channel characteristics for identity authentication. However, most PLA schemes for IIoT require attackers' prior channel information, leading to severe performance degradation when the source of the received signals is unknown in the training stage. Thus, a channel impulse response (CIR)-based PLA scheme named "Hierarchical Variational Auto-Encoder (HVAE)" for IIoT is proposed in this article, aiming at achieving high authentication performance without knowing attackers' prior channel information even when trained on a few data in the complex environment. HVAE consists of an Auto-Encoder (AE) module for CIR characteristics extraction and a Variational Auto-Encoder (VAE) module for improving the representation ability of the CIR characteristic and outputting the authentication results. Besides, a new objective function is constructed in which both the single-peak and the double-peak Gaussian distribution are taken into consideration in the VAE module. Moreover, the simulations are conducted under the static and mobile IIoT scenario, which verify the superiority of the proposed HVAE over three comparison PLA schemes even with a few training data.**

This work was supported in part by the National Natural Science Foundation of China under Grant 61871045 and in part by the Fundamental Research Foundation for the Central Universities under Grant 2022RC15. *(Corresponding author: Xiaodong Xu)*

Rui Meng, Bizhu Wang, and Shujun Han are with the State Key Laboratory of Networking and Switching Technology, Beijing University of Posts and Telecommunications, Beijing 100876, China (e-mail: buptmengrui@bupt.edu.cn; wangbizhu_7@bupt.edu.cn; hanshujun@bupt.edu.cn).

Xiaodong Xu and Ping Zhang are with the State Key Laboratory of Networking and Switching Technology, Beijing University of Posts and Telecommunications, Beijing 100876, China, and also with the Department of Broadband Communication, Peng Cheng Laboratory, Shenzhen 518066, Guangdong, China (e-mail: xuxiaodong@bupt.edu.cn; pzhang@bupt.edu.cn).

Hao Sun was with Peng Cheng Laboratory, Shenzhen 518066, China. He is now with the Department of Applied Mathematics and Theoretical Physics, University of Cambridge, Cambridge CB2 1 TN, UK (e-mail: hsun@pcl.ac.cn).

Shida Xia was with the National Engineering Lab for Mobile Network Technologies, Beijing University of Posts and Telecommunications, Beijing 100876, China. He is now with the China Academy of Information and Communications Technology, Beijing 100191, China (e-mail: xiashida@caict.ac.cn).

## I. Introduction

WITH the deployment of the Internet and information technology, the Industrial Internet of Things (IIoT) has become a growing section to support digital and intelligent industrial production [1]. IIoT connects massive sensing equipment via industrial wireless networks to collect data and make automated decisions [2]. However, the wireless medium makes IIoT face severe threats because of its open nature, mainly spoofing attacks. The spoofing attackers can embezzle other legitimate nodes' identities by tampering with the message’s Media Access Control (MAC) addresses, causing unbearable economic losses and even severe safety accidents [3]. Therefore, identity authentication is an essential guarantee for reliable wireless transmission of IIoT.

Currently, the identity authentication schemes of industrial wireless terminals are mainly based on cryptography, which brings many issues in resource consumption and security [4]-[6]. Specifically, cryptographic authentication relies on negotiating a secret key between the transmitter and the receiver for identity authentication, which suffers once the root key is leaked [4]. Further, the increasing number of access devices in IIoT makes the key distribution management required by cryptographic authentication more difficult, which brings higher computational complexity and latency [7].

By contrast, Physical Layer Authentication (PLA) can achieve lightweight authentication and realize "one secret at a time" by taking advantage of the randomness and time-varying nature of wireless channels [8]-[10]. However, the authentication performance of traditional PLA is determined by the threshold, which is difficult to determine [11]. Although Machine Learning (ML)-based PLA can achieve threshold-free, most existing literature requires attackers' prior channel information, limiting its universality [12]-[14]. Therefore, a lightweight and reliable identity authentication method for IIoT needs to be proposed.

### A. Related Works

The PLA can distinguish legitimate nodes and spoofing nodes by comparing the physical quantity of the received signals and the pilot signal [11], such as Received Signal Strength (RSS) [15] [16], Power Spectral Density (PSD) [17], and Channel Impulse Response (CIR) [18]. The traditional PLA formulates identity authentication into hypothesis testing. The received signal is authenticated as legal if the difference between the received physical quantity and the pilot one does not exceed the threshold. However, the traditional threshold-based PLA can not calculate the optimal threshold value. Instead, it can only identify the received signals by manually traversing the thresholds.

More and more researchers have investigated ML algorithms to achieve threshold-free PLA in recent years [12]-[14]. The edge computing server is physically close to the terminals of IIoT, which is helpful in extracting the channel characteristic parameters. Besides, the server can provide computing resources, and the training stage can be completed at the edge computing server. As a result, the ML-based PLA is suitable for Mobile Edge Computing (MEC)-IIoT [12].

The ML-based PLA can be divided into supervised learning (SL)-based PLA and unsupervised learning (UL)-based PLA. The authors of [12]-[14] proposed several SL-based PLA schemes. Liao *et al.* advocated a deep neural network (DNN)-based authentication scheme in Mobile-Edge Computing (MEC) and verified the authentication performance in IIoT [12]. Xie *et al*. proposed an optimized classification algorithm based on Gaussian support vector machine (SVM) for access authentication in IIoT [13]. Pan *et al.* verified the security of three PLA frameworks based on other ML algorithms, including K-Nearest Neighbors (KNN), Bagged Trees (BT), and Decision Tree (DT) in industrial wireless cyber-physical systems [14]. However, SL-based PLA needs attackers' information for training, limiting its practical application in IIoT.

UL does not need attackers' channel information as labels to guide the training process, consequently more generality. Xia *et al.* considered multiple correlated attributes and proposed a clustering-based authentication algorithm named "Improved System Evolution" (ISE) [19]. However, this scheme is based on statistical channel information, including Channel Gain (CG), channel phase (CP), peak power (PP), and root mean square delay spread (RMS-DS), which represent the coarse-grained information of industrial wireless channels. As a result, there is still room for improvement in industrial wireless security.

Auto-Encoder (AE) and Variational Auto-Encoder (VAE) are powerful tools for unsupervised Deep Learning (DL) in the field of computer image recognition [20]. AE and VAE are based on the standard function approximators (neural networks) and can be trained with gradient descent, attracting more and more researchers from other fields [21]-[23]. [21] designed an end-to-end communication system using the encoder and the decoder of AE to replace the transmitter and the receiver. [22] proposed a network intrusion detection method based on Conditional Variational Auto-Encoder (CVAE). [23] proposed a network intrusion detection architecture based on VAE and Generative Adversarial Networks (GAN) by combining the advantages of adversarial learning and DL. The data sets of the CIR have their characteristics, such as the correlation between subcarriers and time continuity [12], which indicates that AE and VAE can be used to exploit the characteristics of industrial wireless environments.

Due to the long multipath delay and large Doppler shift in IIoT, it is challenging for UL-based PLA (e.g., ISE algorithm [19]) to authenticate the near spoofing nodes. Although SL-based PLA (e.g., EA-DA algorithm [12]) has better authentication performance, it needs attackers' prior channel information, which is difficult to obtain in IIoT. To overcome the above challenges, a promising alternative approach of PLA is proposed in this paper, named as Hierarchical Variational Auto-Encoder (HVAE) scheme, which has better authentication performance and does not require spoofing attackers' prior channel information.

### B. *Contributions*

The proposed HVAE algorithm is applied for learning industrial wireless channels intelligently, thus increasing the authentication rate of legitimate nodes and the detection rate of spoofing nodes.

The main contributions of this paper are described as follows:

1) We convert the authentication problem into a channel difference function and prove that the PLA can be approximated with an arbitrarily small error by a neural network. Then, we propose a CIR-based HVAE-PLA scheme for IIoT to achieve high authentication performance without requiring attackers' prior channel information even when trained on a few data.
2) Unlike the unsupervised clustering-based schemes, we propose the HVAE-PLA scheme based on the unsupervised generative model, including AE and VAE. We propose a hierarchical structure, which includes an AE module for dimension reduction and a VAE module for authentication, which can effectively avoid the over-fitting problem and obtain high authentication when authenticating the CIRs in the IIoT scenario, usually with high dimension.
3) The designed VAE module is equipped with a single-peak Gaussian distribution for CIR reproduction and a revised double-peak Gaussian distribution for classification. The proposed constructed loss function considers both the reconstruction error and the representation quality of CIR, thus improving the security and robustness of PLA in IIoT.
4) The proposed neural network architecture can extract valuable features of high-dimensional CIRs for authentication. The simulations are conducted in two representative real industrial environments, including the static outdoor environment and dynamic indoor environment, where the statistical fading characteristics are Rician fading and Rayleigh fading, respectively. Compared with the ISE algorithm in [19], the simulation results show that the proposed scheme can authenticate the spoofing nodes in all positions in the static data set. The simulation results in the dynamic data set further verify that the proposed method can obtain higher authentication performance than three comparison baselines by 17.18% to 69.3%.

The rest of this paper is organized as follows. In Section II, the system model and problem formulation are described. In Section III, we proposed an HVAE scheme for physical layer authentication. In Section IV, we provided the performance metrics. The simulation results and conclusions are drawn in Section V and Section VI, respectively.

TABLE I
THE LIST OF MAIN PARAMETERS

| Notations | Meanings | Notations | Meanings |
|---|---|---|---|
| $\boldsymbol{X}[t]$ | The CIR vector at time $t$ evaluated by Bob | $\boldsymbol{I}[t]$ | The impulse response of the wireless channel |
| $\boldsymbol{G^{TX}}[t]$ | The antenna gain patterns for the transmitter | $\boldsymbol{G^{RX}}[t]$ | The antenna gain patterns for the receiver |
| $\{P_{A,E}\}$ | The possible positions of Alice and Eves | $\mathcal{D}(\boldsymbol{X}[n])$ | The difference between the $\boldsymbol{X}[n]$ and $\boldsymbol{X}[0]$ |
| $\boldsymbol{W}$ | The weight matrix | $\boldsymbol{B}$ | The bias matrix |
| $\phi$ | The encoder in AE | $\psi$ | The decoder in AE |
| $\phi_1$ | The encoder in the first unit of VAE | $\psi_1$ | The decoder in the first unit of VAE |
| $\phi_2$ | The encoder in the second unit of VAE | $\psi_2$ | The decoder in the second unit of VAE |
| $\beta$ | The learning rate of the neural network | $\rho$ | The maximum number of iterations |
| $\varsigma$ | The batch size | $\alpha$ | The proportion of the legitimate signals |
| $f_{Eve}$ | Eve's attack frequency | $f_{Alice}$ | Alice's signal emission frequency |
| $f_{AB}$ | The carried frequency | $q$ | The number of authenticating signals |
| $\mathcal{L}_1$ | The loss function of AE | $\mathcal{L}_2$ | The loss function of the first unit in VAE |
| $\mathcal{L}_3$ | The loss function of the second unit in VAE | $\mathcal{L}$ | The total loss function |
| $\boldsymbol{Z_2}[n]$ | The mapping of $\boldsymbol{X}[n]$ in the neural network | $\boldsymbol{u_i}$ | The mean vector corresponding to $\boldsymbol{H_i}$ |
| $\boldsymbol{\sigma_i^2}$ | The variance vector corresponding to $\boldsymbol{H_i}$ | $Q_\vartheta$ | The distribution used to approximate $P$ |

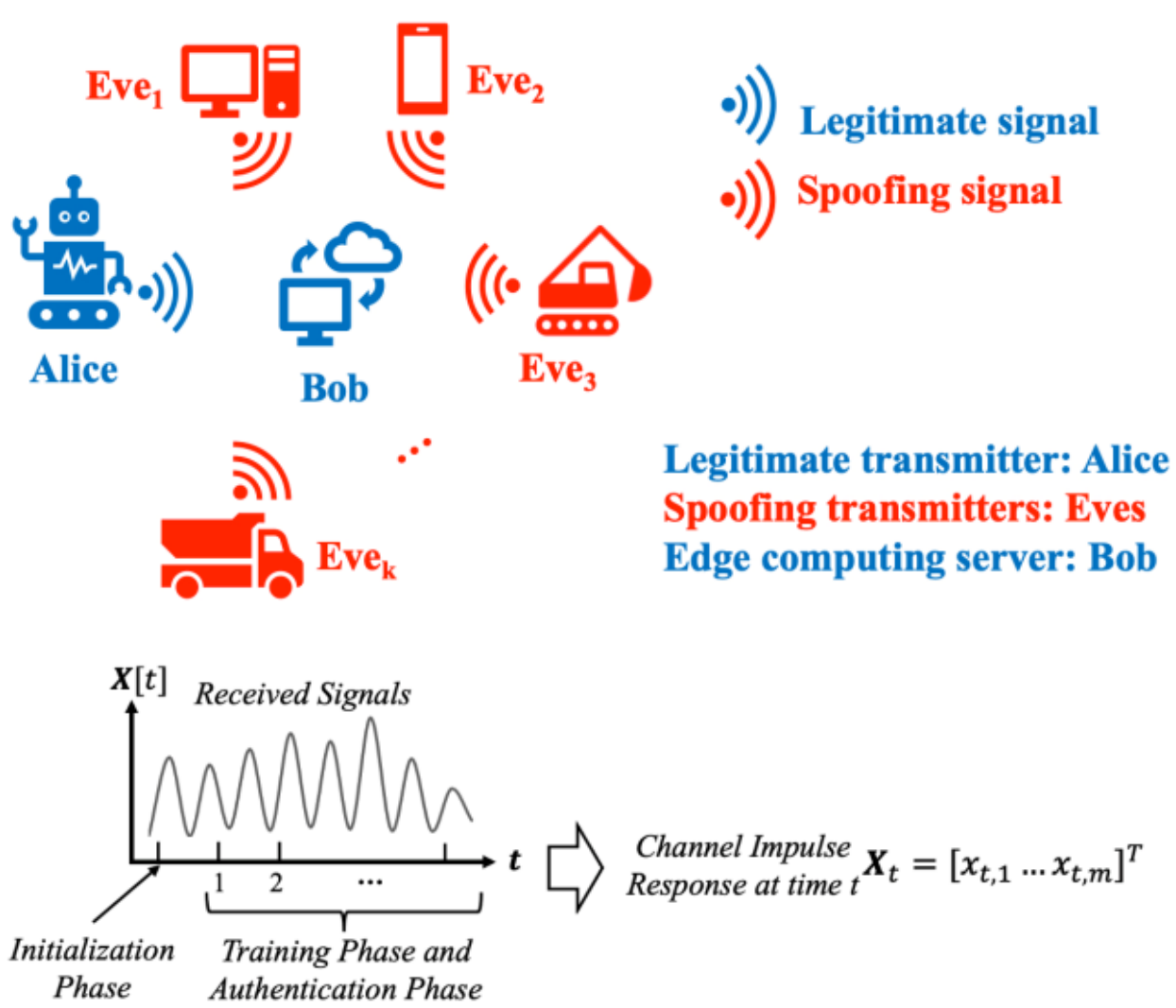

Fig. 1. System model in the Mobile Edge Computing (MEC) industrial wireless communication system, where the legitimate transmitter (Alice) communicates with the edge computing server (Bob) under the attack of multiple spoofing nodes (Eves). Bob's purpose is to identify whether the signals received during the authentication phase come from Alice.

## II. SYSTEM MODEL AND PROBLEM FORMULATION

### A. System Model

Edge computing can provide low-latency and high-reliability services for the IIoT. However, the end users are located at the edge of heterogeneous networks and encounter more serious security problems. Besides, the terminal nodes have limited computing resources and can not follow traditional security protocols [24]. Therefore, our proposed Physical Layer Authentication (PLA) in the IIoT is illustrated in Fig. 1. The legitimate transmitter (Alice) communicates with the edge computing server (Bob) under the attack of multiple spoofing nodes ($\text{Eve}_1$, $\text{Eve}_2$, …, $\text{Eve}_k$). Those Eves pretend to be Alice and send spoofing signals during other time slots to intercept the information. Bob's primary objective is to accurately identify whether the received signals are broadcast by Alice.

The edge computing server is physically close to the terminals, which helps extract the physical channel characteristics of Alice and Eves. Besides, the server can provide computing resources, and the training stage can be completed at the edge computing server. As a result, the PLA based on neural networks is suitable for MEC-IIoT [12]. In this paper, the physical quantity for authentication is Channel Impulse Response (CIR), and the CIR matrix $\boldsymbol{X}[t]$ at time $t$ evaluated by Bob can be derived by (1).

$$\boldsymbol{X}[t] = \boldsymbol{G^{TX}}[t] * \boldsymbol{I}[t] * \boldsymbol{G^{RX}}[t] \quad (1)$$

where $*$ denotes temporal convolution, $\boldsymbol{G^{TX}}[t]$ and $\boldsymbol{G^{RX}}[t]$ are the antenna gain patterns for the transmitter and receiver, respectively, and $\boldsymbol{I}[t]$ is the impulse response of the wireless channel. The channel estimation is not the focus of this paper. The modifications can be made based on (1) for specific application scenarios, which will be described in the exact formula of Section V. The main parameters in this paper are shown in Table I.

### B. Traditional Threshold-Based PLA scheme

The authentication performance of the traditional threshold-based PLA scheme depends on the threshold, and the threshold $\xi$ is obtained by a hypothesis verification $\mathcal{V}$ [12] [16].

$$\mathcal{V} = \begin{cases} \text{diff}\,(\boldsymbol{X}[0], \boldsymbol{X}[n]) < \xi \text{: } \mathbb{H}_0 \\ \text{diff}\,(\boldsymbol{X}[0], \boldsymbol{X}[n]) \geq \xi \text{: } \mathbb{H}_1 \end{cases} \quad (2)$$

where $\mathbb{H}_0$ and $\mathbb{H}_1$ denote *Hypothesis 1* and *Hypothesis 2* respectively.

*Hypothesis 1:* The signal is from Alice;

*Hypothesis 2:* The signal is from Eve.

### C. The Proposed HVAE-based PLA scheme

The proposed threshold-free HVAE-based PLA scheme primarily includes the initialization, training, and authentication phases. The detailed steps are outlined in Algorithm 1. Bob can infer channel features from $\boldsymbol{X}[0]$ as the reference signal and compare the difference between $\boldsymbol{X}[n]$ and $\boldsymbol{X}[0]$. Considering researchers have proposed various methods to detect pilot attacks, such as the beamforming strategy [25], the data analysis algorithms [26], and the upper-layers authentication mechanisms [11] [27], state-of-the-art PLA works focus on achieving high performance given the regular pilot signals. We clarify here that the manuscript's contribution is physical layer-based user-identity authentication given the regular pilot signal. The pilot attack detection is another well-researched area that is out of the scope of this paper.

*Assumption 1:* Bob's observation from wireless channels individualities transmitted from devices in different positions has strong spatial decorrelation, but only when Eves are located more than half a wavelength away from Alice and Bob, the channel-based properties of the received signals at the same receiver can be regarded as entirely irrelevant in time, frequency and space [11], which can be mathematically given as

$$d_{EA} > \frac{c}{2f_{AB}} \tag{3}$$

$$d_{EB} > \frac{c}{2f_{AB}} \tag{4}$$

where $d_{EA}$ represents the distance between Eve and Alice, $d_{EB}$ represents the distance between Eve and Bob, $f_{AB}$ represents the carrier frequency, $c$ is the speed of light. Therefore, the wireless channels of Alice and Eves have strong spatial uniqueness, and it is difficult for Eves to measure, reconstruct and copy [11].

### D. Problem Formulation

Consider the channel model in (1), and the channel is the function of the environment and the antenna gain patterns. Supposed that Bob (the edge server) is fixed, for a given industrial wireless environment, there exists a deterministic mapping function from the positions $P_{A,E}$ of Alice and Eves to the corresponding CIRs $\boldsymbol{X}[t]$ [28].

*Definition 1:* The position-to-channel mapping $\boldsymbol{\Phi}_t$ can be formulated as follows

$$\boldsymbol{\Phi}_t: \{P_{A,E}\} \rightarrow \{\boldsymbol{X}[t]\} \tag{5}$$

where the sets $\{P_{A,E}\}$ denote the possible positions of Alice and Eves, the sets $\{\boldsymbol{X}[t]\}$ denote the CIRs of the corresponding channels.

To achieve authentication, we discussed the existence of the mapping from the CIRs to the positions of Alice and Eves as follows.

*Assumption 2:* The position-to-channel mapping in (5) is bijective.

Although *Assumption 2* can not be proved analytically, based on the uniqueness and time-varying characteristics of wireless channels, the probability that *Assumption 2* holds is very high in industrial wireless environments [29] [30]. As a result, it is reasonable to adopt *Assumption 2* in the IIoT scenario.

**Algorithm 1: Detailed Steps of the Proposed PLA Scheme**

**The Initialization Phase:**
**Step 1:** Alice broadcasts the pilot signal to Bob for access requests over a wireless channel.
**Step 2:** Bob receives the pilot estimated CIR as $\boldsymbol{X}[0]$.
**Step 3:** Bob checks whether $\boldsymbol{X}[0]$ is from Alice to defend against the pilot attacks by utilizing the upper-layers authentication mechanisms.
**Step 4:** If $\boldsymbol{X}[0]$ is from Alice, Bob collects $\boldsymbol{X}[0]$ for training. If not, Bob ignores $\boldsymbol{X}[0]$.

**The Training Phase:**
**Step 5:** Alice sends legitimate signals to Bob over the same wireless channel as in the initialization phase.
**Step 6:** All Eves pretend to be Alice and send spoofing signals to Bob during other time slots to intercept the information. Bob utilizes the upper-layers authentication mechanisms to detect the pilot attacks in the training phase.
**Step 7:** Bob uses **Algorithm 2** to train neural networks through $\boldsymbol{X}[0]$, the corresponding tag of $\boldsymbol{X}[0]$ and the received signals $\boldsymbol{X}[1] \sim \boldsymbol{X}[p]$ in the training phase.

**The Authentication Phase:**
**Step 8:** Alice and Eves send signals to Bob for authentication.
**Step 9:** Bob receives multiple signals $\boldsymbol{X}[p+1] \sim \boldsymbol{X}[p+q]$ and compares them with $\boldsymbol{X}[0]$ through (39) after the training phase.
**Step 10:** After **Step 9**, Bob can distinguish whether the signals $\boldsymbol{X}[p+1] \sim \boldsymbol{X}[p+q]$ are from Alice or not.
**Step 11:** Return to **The Initialization Phase** for authentication of subsequent signals.

*Definition 2:* Under *Assumption 2*, the mapping from CIRs to the positions of Alice and Eves, $\boldsymbol{\Phi}_t^{-1}$, exists, which can be formulated as

$$\boldsymbol{\Phi}_t^{-1}: \{\boldsymbol{X}[t]\} \rightarrow \{P_{A,E}\} \tag{6}$$

where $\boldsymbol{\Phi}_t^{-1}$ denotes the inverse mapping of $\boldsymbol{\Phi}_t$.

The pilot attacks in the initialization phase can be detected, so we can obtain the mapping as $\boldsymbol{X}[0] \rightarrow P_A$. $P_A$ denotes that $\boldsymbol{X}[0]$ is from Alice. Under *Assumption 1*, the authentication phase is to compare the channel differences, which can be formulated as

$$\mathcal{D}(\boldsymbol{X}[t_A]) < \mathcal{D}(\boldsymbol{X}[t_E]) \tag{7}$$

where the channel difference function $\mathcal{D}$ can be formulated as

$$\mathcal{D}(\boldsymbol{X}[n]) = \text{diff}\,(\boldsymbol{X}[0], \boldsymbol{X}[n]) \tag{8}$$

Therefore, $\mathcal{D}(\boldsymbol{X}[t_A])$ indicates the difference between $\boldsymbol{X}[0]$ and the signals from Alice, and $\mathcal{D}(\boldsymbol{X}[t_E])$ indicates the difference between $\boldsymbol{X}[0]$ with the signals from Eves. In the authentication phase, there are $q$ CIRs, including $\boldsymbol{X}[p+1]$, $\boldsymbol{X}[p+2]$, ..., $\boldsymbol{X}[p+q]$. The channel differences can be denoted as $\mathcal{D}(\boldsymbol{X}[p+1])$, $\mathcal{D}(\boldsymbol{X}[p+2])$, ..., $\mathcal{D}(\boldsymbol{X}[p+q])$. Sorting them according to a numerical value, we can obtain

$$\mathcal{D}(\boldsymbol{X}[\dot{1}]) < \mathcal{D}(\boldsymbol{X}[\dot{2}]) < \cdots < \mathcal{D}(\boldsymbol{X}[\dot{q}]) \quad (9)$$

Based on (7), the signals corresponding to the first $\alpha q$ values are authenticated as legitimate, and the others are authenticated as illegal, which can be written as the mapping (10) and (11)

$$\{\mathcal{D}(\boldsymbol{X}[\dot{1}]), \mathcal{D}(\boldsymbol{X}[\dot{2}]), \ldots, \mathcal{D}(\boldsymbol{X}[\dot{\alpha q}]\} \rightarrow \{P_A\} \quad (10)$$

$$\{\mathcal{D}(\boldsymbol{X}[\alpha q \dot{+} 1]), \mathcal{D}(\boldsymbol{X}[\alpha q \dot{+} 2]), \ldots, \mathcal{D}(\boldsymbol{X}[\dot{q}]\} \rightarrow \{P_E\} \quad (11)$$

where $\alpha$ denotes the proportion of the legitimate signals in the authentication phase. $\alpha$ is determined by Alice's signal emission frequency $f_{Alice}$ and Eve's attack frequency $f_{Eve}$, which can be written as follows

$$\alpha = \frac{f_{Alice}}{f_{Alice} + f_{Eve}} \quad (12)$$

*Proposition 1:* Based on (8), (10), and (11), the identification and authentication process can be written as the channel-to-position mapping $\boldsymbol{\Phi}_{\mathrm{I\&A}}$

$$\boldsymbol{\Phi}_{\mathrm{I\&A}}: \{\boldsymbol{X}[n]\} \rightarrow \{P_{A,E}\}, n = [p+1, p+q] \quad (13)$$

Therefore, the focus of this article is to find the function $\mathcal{D}(\boldsymbol{X}[n])$ to measure the difference between the received signals $\boldsymbol{X}[n]$ and the reference signal vector $\boldsymbol{X}[0]$ so that we can use mapping $\boldsymbol{\Phi}_{\mathrm{I\&A}}$ to achieve PLA.

### *E. Deep Learning for Physical Layer Authentication*

*Proposition 1* proves the existence of the channel-to-position mapping function in the authentication phase. However, the function can not be described by known mathematical models. Therefore, we use a deep learning algorithm to obtain $\mathcal{D}(\boldsymbol{X}[n])$. Based on the universal approximation theorem [31], we can obtain *Proposition 2.*

*Proposition 2:* For any given small error $\epsilon > 0$, there always exists a positive constant $N$ large enough such that

$$\sup_{\boldsymbol{X} \in \mathbb{C}} \left|\left|\mathrm{NET}_N(\boldsymbol{X}, \boldsymbol{\Omega}) - \boldsymbol{\Phi}_{\mathrm{I\&A}}(\boldsymbol{X})\right|\right| \leq \epsilon \quad (14)$$

where $\mathrm{NET}_N(\boldsymbol{X}, \boldsymbol{\Omega})$ is the authentication results based on a neural network with $\boldsymbol{X}$, $\boldsymbol{\Omega}$, and $N$ denoting the input CIRs, network parameters, and the number of hidden units, respectively.

*Proof:* (i) Since CIR $\boldsymbol{X}$ is bounded and closed, $\mathbb{C}$ is a compact set; (ii) Since $\boldsymbol{\Phi}_t^{-1}$, mapping (8), (10), and (11) are continuous mappings, we know that for $\forall \boldsymbol{X} \in \mathbb{C}$ such that $\boldsymbol{\Phi}_{\mathrm{I\&A}}(\boldsymbol{X})$ is a continuous function. Based on (i), (ii), and the universal approximation theorem [31], *Proposition 2* is proved.

According to *proposition 2*, the channel-to-position mapping function can be approximated with an arbitrarily small error by a neural network. Therefore, we proposed a PLA scheme based on HVAE to authenticate legitimate signals and spoofing signals through dimension reduction and feature extraction.

## III. Proposed Physical Layer Authentication Scheme Based on Hierarchical Variational Auto-Encoder

Deep learning shows advantages in automatic learning of wireless channel features from received signals. Auto-Encoder (AE) has two parts: the encoder and the decoder. The encoder can reduce the dimension of the input wireless signal vectors to obtain the hidden space with a lower dimension. The decoder can generate a new space with the same dimension as the input signal vectors. Variational Auto-Encoder (VAE) [20] maps each training sample to a distribution instead of to a unique hidden vector to overcome the over-fitting issue in AE. The single-peak distribution is intuitively chosen to re-generate the training sample.

The successful generation models usually use only one layer of hidden variables [32] [33], and those models using multiple layers only show moderate performance growth in quantitative indicators (such as log-likelihood) [34] [35]. Our architecture used shallow neural networks to separate the dimension reduction and classification process.

However, when using more hidden variables in AE and VAE, representing CIRs would be sufficient to generate CIRs with high correlation. However, the disadvantage is that the accuracy of unsupervised classification is limited. Using fewer hidden variables makes it convenient to obtain better classification performance, but representing CIRs would be insufficient to generate CIRs with high correlation [20].

*Remark 1:* To combine the pros and avoid the cons mentioned above, we introduced a Hierarchical Variational Auto-Encoder (HVAE) algorithm based on AE and VAE to obtain better authentication performance. The designed structure is illustrated in Fig. 2. The AE here, which has three-layer neural networks with full connection as the encoder $\phi$ and three-layer neural networks with full connection as the decoder $\psi$, acts as a low-level feature extractor. AE encodes the input space $X$ with 8,188 dimensions as a hidden space $H$ with $h$ dimensions and decodes the hidden space $\dot{H}$ with $h$ dimensions as new samples $\dot{X}$. In contrast, the VAE here serves as a high-level classifier with two hidden units $Z_1$ and $Z_2$. $Z_1$ and $Z_2$ are both fully connected. $Z_1$ (including encoder $\phi_1$ and decoder $\psi_1$) follows the ordinary VAE to use a simple Gaussian prior $\mathcal{N}(\mu, \sigma^2)$, which can store available information for VAE-reproduction, and $Z_2$ (including encoder $\phi_2$ and decoder $\psi_2$) is equipped with a revised double-peak Gaussian prior $\alpha\mathcal{N}(-\mu, \sigma^2) + (1-\alpha)\mathcal{N}(\mu, \sigma^2)$ to encourage separation, where $\alpha$ denotes a weighting factor.

Instead of employing VAE as a generative method that only focuses on the reconstruction error and the hidden layer representation quality, the classification accuracy also matters to the proposed scheme. Therefore, the single-peak distribution involved in the plain VAE is replaced with the double-peak distribution, which maps the training sample to a distribution instead of a vector (as in plain VAE) and matches the goal as user-identity classification simultaneously. The detailed algorithm is shown in Algorithm 2, where $\boldsymbol{W}$ and $\boldsymbol{B}$ indicate weight and bias, respectively. $\beta$ denotes the learning rate of the neural network. $\rho$ denotes the maximum number of iterations. $\varsigma$ denotes the batch size. In the following part of this section, our scheme includes four stages which are pre-process, dimension reduction process, further dimension reduction process, and classification process.

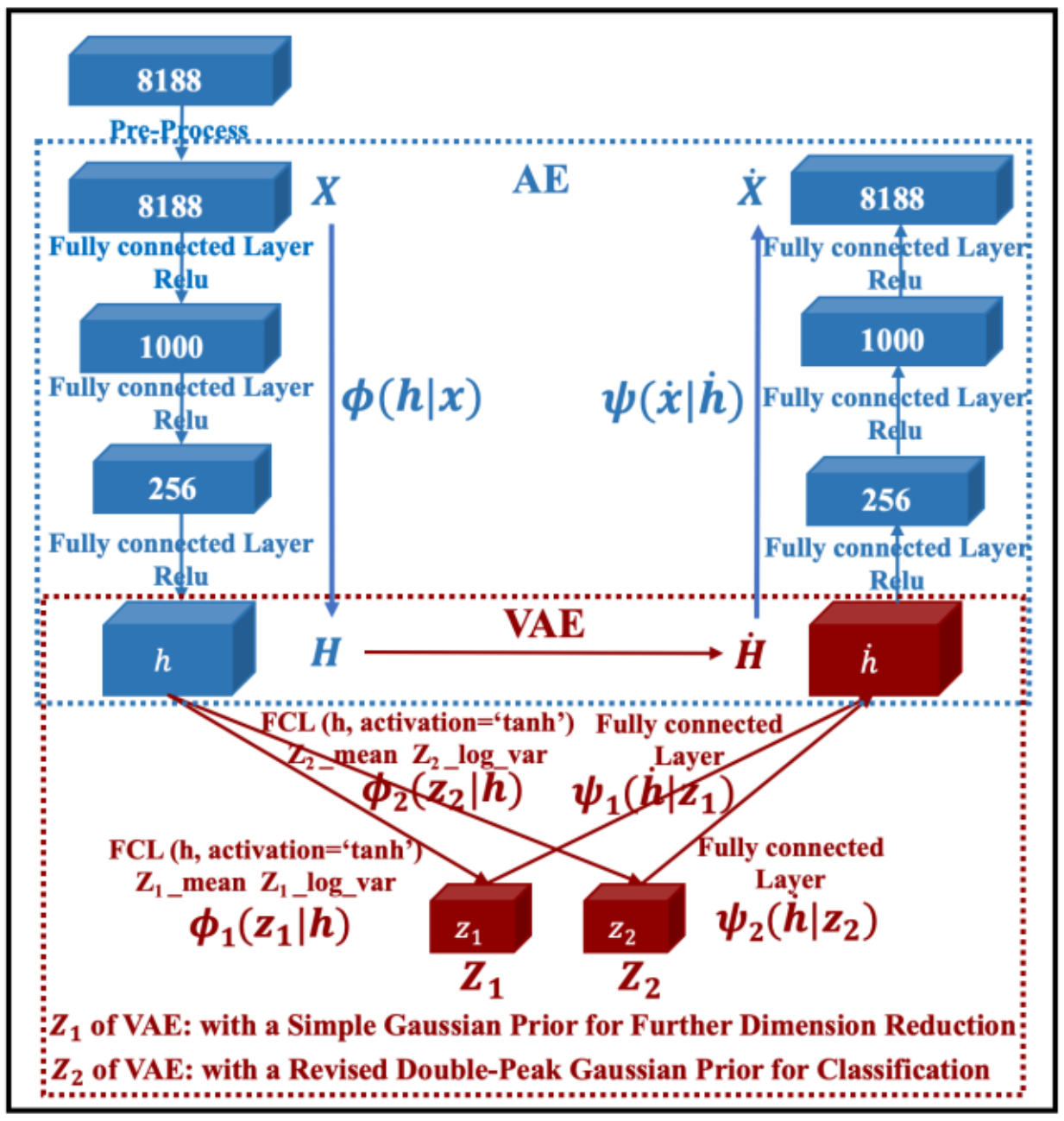

Fig. 2. The designed Hierarchical Variational Auto-Encoder (HVAE) with a hierarchical structure based on AE and VAE, where "8,188" corresponds to the dimension of the CIR in Section V.A.

### A. *Pre-Process*

Channel Impulse Response (CIR) is a multidimensional complex vector $\boldsymbol{X_C}$, and the normalization is shown in (15) to obtain normalized CIR

$$\boldsymbol{X} = (\boldsymbol{X_C} - E[\boldsymbol{X_C}])/D[\boldsymbol{X_C}] \tag{15}$$

where $E[\boldsymbol{X_C}]$ and $D[\boldsymbol{X_C}]$ denote the average and variance of samples $\boldsymbol{X_C}$, respectively.

### B. *AE for Dimension Reduction*

AE can be used for the dimension reduction of CIRs by selectively retaining some existing features or generating fewer neat features based on the combination of old components. The encoder and the decoder in AE can be defined as transitions $\phi$ and $\psi$, respectively, such that

$$\phi\colon X \to H \tag{16}$$

$$\psi\colon \dot{H} \to \dot{X} \tag{17}$$

where $H$ denotes the hidden space with $h$ dimensions, $\dot{H}$ is obtained by VAE including encoders ($\phi_1$ and $\phi_2$) and decoders ($\psi_1$ and $\psi_2$), and $\dot{X}$ denotes the generated new CIRs.

*Proposition 3:* We can obtain the objection function $\mathcal{L}_1$ of AE as

$$\mathcal{L}_1 = \| \boldsymbol{X} - \psi\,[(\phi_1 \circ \psi_1) + (\phi_2 \circ \psi_2)\,](\phi \boldsymbol{X})\|^2 \tag{18}$$

where $\| \ \|^2$ denotes the reconstruction error of AE.

Fig. 2 illustrates the proposed neural network, including an input layer, multiple hidden layers and an output layer. Each layer includes multiple neurons, and the input of each neuron is the sum of the output of the last neuron multiplied by the corresponding coefficient. The output of each neuron is obtained after an activation function activates its input. Activation functions include Sigmoid function, Tanh function, and Rectified Linear Unit (ReLU) function as

$$\sigma_{\text{Sigmoid}}(\boldsymbol{z}) = \frac{1}{1 + e^{-\boldsymbol{z}}} \tag{19}$$

$$\sigma_{\text{Tanh}}(\boldsymbol{z}) = \frac{e^{\boldsymbol{z}} - e^{-\boldsymbol{z}}}{e^{\boldsymbol{z}} + e^{-\boldsymbol{z}}} \tag{20}$$

$$\sigma_{\text{ReLU}}(\boldsymbol{z}) = \max(\boldsymbol{0}, \boldsymbol{z}) \tag{21}$$

where $\boldsymbol{z}$ denotes a vector. Sigmoid function has its advantages (it can map the output of neural network between 0 and 1, which is used to denote the classification probability). However, the disadvantage is that the mean value of the output is not equal to 0, so it is not suitable for hidden layer activation of multilayer neural networks. Besides, when the function value is large, the derivative equals 0, which causes the vanishing gradient problem. ReLU function can avoid the vanishing gradient problem and is usually used as the activation function of deep learning algorithms. The mean value of the output of Tanh function is equal to 0, which is suitable for multi-layer activation. As a result, we used ReLU function and Tanh function as the activation function in the AE and VAE, respectively.

The output of each layer of neural network is denoted as

$$\boldsymbol{Z_l} = f^{l}_{\text{activation}}(\boldsymbol{W_l Z_{l-1}} + \boldsymbol{B_l}) \tag{22}$$

where $\boldsymbol{Z_l}$ denotes the output of the $l$-th layer network, $\boldsymbol{W_l}$ represents the weight matrix of the $l$-th layer and the pervious layer, $\boldsymbol{B_l}$ denotes the bias matrix.

We use back propagation and gradient descent to optimize $\boldsymbol{W_l}$ and $\boldsymbol{B_l}$ [27] and minimize the reconstruction error $\mathcal{L}_1$ in (18). The parameters in the $l$-th layer network can be formulated as

$$\nabla \boldsymbol{W_l} = \frac{1}{\zeta} \nabla \boldsymbol{Z_l} * \left(f^{l}_{\text{activation}}\right)' |_{\boldsymbol{Z_l}} \cdot \boldsymbol{Z}^{T}_{\boldsymbol{l-1}} \tag{23}$$

$$\nabla \boldsymbol{B_l} = \frac{1}{\zeta} \nabla \boldsymbol{Z_l} * \left(f^{l}_{\text{activation}}\right)' |_{\boldsymbol{Z_l}} \cdot \boldsymbol{e_1} \tag{24}$$

**Algorithm 2: The Threshold-Free Hierarchical Variational Auto-Encoder Authentication Algorithm**

Require $\boldsymbol{X_c}, \eta, \beta, \rho, \varsigma$
Obtain normalized CIR $\boldsymbol{X}$ by (11)
Initialize $\boldsymbol{W}$ and $\boldsymbol{B}$
Initialize $\boldsymbol{H} = \boldsymbol{\phi}(\boldsymbol{X},\ \boldsymbol{W}^{[\boldsymbol{\phi}]}, \boldsymbol{B}^{[\boldsymbol{\phi}]})$
Initialize $\boldsymbol{Z_1} = \boldsymbol{\phi_1}(\boldsymbol{H}, \boldsymbol{W}^{[\boldsymbol{\phi_1}]}, \boldsymbol{B}^{[\boldsymbol{\phi_1}]})$
Initialize $\boldsymbol{Z_2} = \boldsymbol{\phi_2}(\boldsymbol{H}, \boldsymbol{W}^{[\boldsymbol{\phi_2}]}, \boldsymbol{B}^{[\boldsymbol{\phi_2}]})$
Initialize $\dot{\boldsymbol{H}} = \boldsymbol{\psi_1}(\boldsymbol{Z_1}, \boldsymbol{W}^{[\boldsymbol{\psi_1}]}, \boldsymbol{B}^{[\boldsymbol{\psi_1}]}) + \boldsymbol{\psi_2}(\boldsymbol{Z_2}, \boldsymbol{W}^{[\boldsymbol{\psi_2}]}, \boldsymbol{B}^{[\boldsymbol{\psi_2}]})$
Initialize $\dot{\boldsymbol{X}} = \boldsymbol{\psi}(\dot{\boldsymbol{H}}, \boldsymbol{W}^{[\boldsymbol{\psi}]}, \boldsymbol{B}^{[\boldsymbol{\psi}]})$
**for** $i$ in $\boldsymbol{\eta}$ **do**
  Update $\boldsymbol{W}$ and $\boldsymbol{B}$ by (28) and (29)
  Calculate $\mathcal{L}_1$ by (18)
  Calculate $\mathcal{L}_2$ by (33)
  Calculate $\mathcal{L}_3$ by (35)
  Calculate $\mathcal{L} = \mathcal{L}_1 + \mathcal{L}_2 + \mathcal{L}_3$ by (38)
**end for**
Obtain $\boldsymbol{Z_2}(\boldsymbol{n})$
Calculate diff $(\boldsymbol{X}[\boldsymbol{0}], \boldsymbol{X}[\boldsymbol{n}])$ by (39)
Obtain authentication results

$$\nabla \boldsymbol{Z_l} = \boldsymbol{W_{l+1}} \cdot \left(\nabla \boldsymbol{Z_{l+1}} * \left(f^{l}_{\text{activation}}\right)' |_{\boldsymbol{Z_{l+1}}}\right) \tag{25}$$

where $\boldsymbol{e_1} = (1, 1, \ldots, 1)$, $\zeta$ denotes the number of "1" in $\boldsymbol{e_1}$. The gradient descent is shown as

$$\nabla \boldsymbol{W}_l^+ = \varpi^+ \nabla \boldsymbol{W}_l^+ + (1 - \varpi^+) \nabla \boldsymbol{W_l} \tag{26}$$

$$\nabla \boldsymbol{B}_l^+ = \varpi^+ \nabla \boldsymbol{B}_l^+ + (1 - \varpi^+) \nabla \boldsymbol{B_l} \tag{27}$$

where $\varpi^+$ denotes the coefficients of gradient descent. The parameters are updated as

$$\boldsymbol{W_l} \leftarrow \boldsymbol{W_l} - \beta \nabla \boldsymbol{W}_l^+ \tag{28}$$

$$\boldsymbol{B_l} \leftarrow \boldsymbol{B_l} - \beta \nabla \boldsymbol{B}_l^+ \tag{29}$$

where $\beta$ denotes the learning rate of the neural network.

### C. $Z_1$ of VAE for Further Dimension Reduction

The input of $Z_1$ is encoded as a probability distribution in hidden space instead of being encoded as a single point like (16) and (17) in AE.

The architecture of $H \to Z_1 \to \dot{H}$ follows the ordinary VAE, where each sample $H_i$ corresponds to a mean vector $u_i$ and a variance vector $\sigma_i^2$, so there is an exclusive normal distribution matching $H_i$. It can adjust the variance $\sigma_i^2$ using this exclusive normal distribution, and random sampling from this distribution will generate samples $\dot{H}_i$ similar to input samples $H_i$ [36].

We approximate $P(Z_1|H)$ with another distribution $Q_\vartheta(Z_1|H)$ by variational inference and use Kullback-Leibler divergence (KL divergence) to measure the distance between $P(Z_1|H)$ and $Q_\vartheta(Z_1|H)$

$$KL(Q_\vartheta(Z_1|H)||P(Z_1|H))$$
$$= KL(Q_\vartheta(Z_1|H)||P(Z_1)) - \mathbb{E}_{Z_1 \sim Q_\vartheta(Z_1|H)}[\ln P(H|Z_1)] + \log P(H) \tag{30}$$

*Proof:* See Appendix A.

$$KL(Q_\vartheta(Z_1|H)||\mathcal{N}(0, I))$$
$$= \frac{1}{2}(-\log \sigma^2 + \sigma^2 + \mu^2 - 1) \tag{31}$$

*Proof:* See Appendix B.

Then we sample by $P(Z_1|H)$ for subsequent calculation, and the reparameterization trick is to sample $Z_{1,i}$ from $\mathcal{N}(\mu, \sigma^2)$, that is, sample $\epsilon$ from $\mathcal{N}(0, I)$, and then calculate it by (32)

$$Z_{1,i} = \mu + \epsilon \times \sigma \tag{32}$$

*Proposition 4:* The objective function $\mathcal{L}_2$ of $H \to Z_1 \to \dot{H}$ is related to the reconstruction error $\mathbb{E}_{Z_1 \sim Q_\vartheta(Z_1|H)}[\log P(H|Z_1)]$ and the KL divergence (31), and we can obtain $\mathcal{L}_2$ as

$$\mathcal{L}_2 = \frac{1}{n}\sum_{i=1}^{n}\left[\log p(H|Z_1) + \frac{1}{2}\sum_{i=1}^{d}(-\log \sigma^2 + \sigma^2 + \mu^2 - 1)\right] \tag{33}$$

where $n$ is the number of CIRs, and $d$ is the dimension of normal distribution.

### D. $Z_2$ of VAE for Classification

Vanilla VAE takes a Gaussian prior in its hidden space to facilitate computation and enables generating CIRs by adding perturbations into the hidden variables before putting them into the decoder networks. In order to obtain better classification performance, we designed a double-peak Gaussian prior in $Z_2$ of VAE.

Denoting the prior of $Z_2$ as $P(Z_2)$, we use

$$P(Z_2) \sim \alpha\mathcal{N}(-\mu, \sigma^2) + (1 - \alpha)\mathcal{N}(\mu, \sigma^2) \tag{34}$$

where $\alpha$ is a weighting factor.

*Proposition 5:* The objection function $\mathcal{L}_3$ of $H \to Z_2 \to \dot{H}$ can be denoted as (35).

$$\mathcal{L}_3 = KL(N(\mu, \sigma^2)||\alpha N(-m, s^2) + (1 - \alpha)N(m, s^2)) \tag{35}$$

Calculation of the KL-divergence of multi-peak-Gaussian cases calls for lots of work. Here we proposed an approximation of the scaling method when $\alpha = \frac{1}{2}$ as (36), a more straightforward approach. An exact calculation is also included in (37).

$$KL(N(\mu, \sigma^2)||\frac{1}{2}N(-m, s^2) + \frac{1}{2}N(m, s^2)$$

$$\leq -\frac{1}{2}\log 2\left(s^2+\log\sigma^2-\frac{1}{2}(\mu-m)^2-\frac{1}{2}(\mu+m)^2-\sigma^2\right) \tag{36}$$

*Proof:* See Appendix C.

With an approximation of $Erfc(x) \approx 1 - tanh(1.19x)$

$$KL(N(\mu,\sigma^2)||\frac{1}{2}N(-m,s^2)+\frac{1}{2}N(m,s^2))$$
$$\approx \log\frac{2s}{\sigma}+\frac{(m-\mu)^2+\sigma^2-s^2}{2s^2}$$
$$+\frac{2m\left\{-\sigma e^{-\frac{\mu^2}{2\sigma^2}}+\sqrt{\frac{\pi}{2}}\mu\left[1-\tanh\left(1.19\frac{\mu}{\sqrt{2}\sigma}\right)\right]\right\}}{s^2} \tag{37}$$

*Proof:* See Appendix D.

The total loss $\mathcal{L}$ can be denoted as

$$\mathcal{L}=\mathcal{L}_1+\mathcal{L}_2+\mathcal{L}_3 \tag{38}$$

We can minimize $\mathcal{L}$ through gradient descent. When the training reaches convergence, the neural network can generate new samples $\dot{X}$ similar to the real signal vectors $X$, which indicates that the neural network has learned the intrinsic characteristics of the wireless signals to use $Z_2[n]$ for classification. $\boldsymbol{Z_2}[n]$ is the mapping of $\boldsymbol{X}[n]$. The channel difference can be formulated as

$$\mathcal{D}(\boldsymbol{X}[n]) = \text{diff}\,(\boldsymbol{X}[0],\boldsymbol{X}[n]) = d_{Euc}(\boldsymbol{Z_2}[0],\boldsymbol{Z_2}[n])$$
$$=\|\boldsymbol{Z_2}[n]-\boldsymbol{Z_2}[0]\|_2^2 \tag{39}$$

where $d_{Euc}$ denotes the Euclidean distance. Minimizing the loss function $\mathcal{L}$ can automatically drive $\mathcal{D}(\boldsymbol{X}[n])$ to be a good classifier.

## IV. Performance Metrics

In this section, $F_1$ Measure and the loss function in (38) are used as the performance metrics. We further analyzed the computational complexity of the proposed authentication scheme.

### A. $F_1$ Measure

$F_1$ Measure is a common evaluation index in the binary classification model of machine learning [37]. In machine-learning-based PLA works [19], $F_1$ is used to evaluate the authentication performance of PLA schemes. The confusion matrix is used to visualize the performance of unsupervised classification algorithms. Each column represents the predicted value, and each row represents the actual category. As illustrated in Table II, we defined the authentication results in the confusion matrix.

True Legitimate (TL) represents the signal of Alice is ideally identified as legality. False Legitimate (FL) means the signal of Eve is recognized as a legality by mistake. False Attack (FA) represents the signal of Alice is mistakenly identified as spoofing. True Attack (TA) means the signal of Eve is rightly recognized as spoofing. Specifically, we concentrate on the correct authentication rate $P_{ca}$ and non-omission authentication rate $P_{noa}$, which can be calculated as

$$P_{ca}=\frac{TL}{TL+FL} \tag{40}$$

$$P_{noa}=\frac{TL}{TL+FA} \tag{41}$$

where $P_{ca}$ denotes the ratio of the signals transmitted from Alice among all the signals authenticated as Alice. $P_{noa}$ represents the ratio of the signals certified as Alice to all the signals sent by Alice. $P_{ca}$ and $P_{noa}$ refer to *precision/confidence* and *recall/sensitivity*, respectively [37]. *Recall* or *sensitivity* (called in Psychology) is the proportion of true positive cases that are correctly predicted to be positive. On the contrary, *accuracy* or *confidence* (called in data mining) indicates the proportion of correct true positive cases among the predicted positive cases. To consider the evaluation criteria of $P_{ca}$ and $P_{noa}$, we utilize $F_\beta$ Measure as (42) to measure authentication performance.

TABLE II
THE CONFUSION MATRIX OF ALICE AND EVE

| Actual \ Predicted | Alice | Eve |
|---|---|---|
| Alice | True Legitimate (TL) | False Attack (FA) |
| Eve | False Legitimate (FL) | True Attack (TA) |

$$F_\beta\text{ Measure}=\frac{(\beta^2+1)\cdot P_{ca}\cdot P_{noa}}{\beta^2\cdot P_{ca}+P_{noa}} \tag{42}$$

When $\beta>1$, the weight of $P_{noa}$ is higher than that of $P_{ca}$; When $\beta<1$, the weight of $P_{ca}$ is higher than that of $P_{noa}$. When $\beta=1$, we can obtain (43)

$$F_1\text{ Measure}=\frac{2*P_{ca}\cdot P_{noa}}{P_{ca}+P_{noa}} \tag{43}$$

The larger $F_1$ Measure, the higher the accuracy of our model, which indicates the more heightened the accurateness of authentication.

### B. Loss Function

Minimizing the loss function is the optimization goal of the model in the training process through gradient descent. The loss function of the HVAE algorithm is based on reconstruction loss and KL-divergence, which are expressed in (18), (33), (35) and (38).

### C. Complexity Analysis

This section focuses on the computational resources in the authentication phase. The HVAE-based PLA scheme needs to perform (20), (21), and (38), and the total number of layers in the neural network is 8, so the computational complexity of the mathematical operation can be approximately denoted as $O(max(n^0 \times n^1, n^1 \times n^2, \ldots, n^7 \times n^8))$, where $n^l$ denotes the number of neurons in the $l$th layer in HVAE.

## V. Simulations

### A. The Description of Industrial Data Set

In order to verify the authentication performance of our proposed authentication scheme in the real industrial wireless environment, we chose the data sets of Channel Impulse Response (CIR) collected by the National Institute of Standards and Technology (NIST) [38]. The NIST conducted the CIR measurements at four facilities representing different classes of industrial environments. The selected sites included:

- the NIST Open Area test site (OATS);
- a typical multi-acre gearbox assembly plant in the automotive factory (AF);
- a small machinery shop mainly engaged in metal processing located on the NIST campus in Gaithersburg;
- the steam generation plant at the NIST campus in boulder.

The OATS and the AF are considered in our simulations to extract the intrinsic features of industrial wireless channels in a static environment and a mobile environment, respectively. The OATS factory is located on the NIST campus in Boulder, Colorado. The AF is an indoor transmission assembly factory covering more than acres. The floor size of the factory exceeds 400 m × 400 m. The ceiling is about 12 meters (40 feet) high. The enclosed space stores small parts and tools in factory inventory.

In the OATS and the AF, the measurement and propagation characteristics of the wireless channel can be obtained by a channel sounder. The channel sounder collects impulse response data of spatial and temporal behavior of the wireless channel. The impulse response measures the electromagnetic environment of the channel by capturing the reflection, diffraction, and scattering phenomena between the transmitter and the receiver.

The NIST channel sounder system for the measurements of the wireless environment is a positive-negative (PN) sequence correlation system [39]. It consists of a single transmitter (TX) and a single receiver (RX) synchronized with two rubidium clocks. The sounder can gather data for carrier frequencies between 65 MHz to 6GHz with an operational bandwidth of 200MHz and a Vector Signal Transceiver (VST). The transmitter can generate a 2,047b PN code sequence, which modulates a Binary Phased-Shift Keying (BPSK) signal. The receiver evaluates the PN sequence with four oversampling rates, so every estimated CIR is expressed as a complex vector with a dimension of 8,188 (a 2,047-point ideal PN sequence times four samples/symbol equals 8,188). The transmitter channel sounder repetitively transmits the $PN_{ideal}(\tau)$ . After propagating through the Tx channel sounder non-ideal hardware responses $x^{TX}(\tau)$, the industrial environment $x(\tau)$, and the Rx channel sounder non-ideal channel sounder hardware responses $x^{RX}(\tau)$, the signal is measured by the Rx channel sounder as $S_{meas}(\tau)$, as seen in (44).

$$S_{meas}(\tau) = PN_{ideal}(\tau) * x^{TX}(\tau) * g^{TX}(\tau) * x(\tau) * g^{RX}(\tau) * x^{RX}(\tau) \tag{44}$$

To correct the linear distortion introduced by non-ideal system hardware and estimate CIR more accurately, the calibrated and filtered CIR $X(\tau)$ is calculated through "back-to-back" measurement [39] as

$$X(\tau) = \mathcal{F}^{-1}\left\{\mathcal{F}[w(\tau)] \times \frac{\mathcal{F}[S_{meas}(\tau)]}{\mathcal{F}[S^{B2B}_{meas}(\tau)] / \mathcal{F}[A]}\right\} \tag{45}$$

where $\mathcal{F}^{-1}$ denotes the inverse Fourier transform, $S^{B2B}_{meas}(\tau)$ denotes the back-to-back-measured signal [39], $A$ is a function of frequency, and $w(\tau)$ is a bandpass filter as

$$w(\tau) = \mathcal{F}^{-1}\left\{\mathcal{F}\left[\frac{PN_{ideal}(\tau)}{2 \times \sqrt{N}}\right] \times \mathcal{F}^{*}\left[\frac{PN_{ideal}(\tau)}{2 \times \sqrt{N}}\right]\right\} \tag{46}$$

where $PN_{ideal}(\tau)$ denotes the oversampled PN sequence utilized by the transmitter where $N$ represents the maximum length sequence order and parted by $2 \times \sqrt{N}$ to obtain unity gain.

### B. Simulation Results and Analysis

#### 1) Simulations with the OATS:

The OATS is a typical Rician fading channel with a strong line of sight (LOS) component. The edge computing server (Bob) receives signals sent by the transmitter from 45 nodes in turn. Fig. 3 illustrates 45 transmitting nodes in OATS where the distances of different nods are 0.5 m, 1 m, and 2 m, and node 23 is the legal transmitting node (Alice) while the others are spoofing nodes. The locations of Alice and Bob are (0,11.5) and (0,0), respectively. Each transmitting node has 400 CIR records, and the receiver estimates the CIRs after the transmitter stops at specific sites. Table III shows the simulation parameters in the OATS. In order to verify the authentication performance of our proposed scheme, we selected the first 40 CIRs of 400 CIRs for training and testing at each transmitting node. The OATS data set for simulation do not include the multi-antenna CIRs data. Therefore, the pilot attacks during the initialization and training phases can be detected by the upper-layers authentication mechanisms.

In the scenario of OATS, we used 30 CIR samples per node for training and 10 CIR samples per node for testing. There are 45 transmission nodes, so we trained 1,350 (30 × 45) CIRs with Auto-Encoder (AE), Hierarchical Variational Auto-Encoder (HVAE), and the Improved System Evolution (ISE) [19] respectively, and then tested the remaining 10 CIRs of each node. Because our authentication process is based on unsupervised learning, we give the unique label to the first CIR transmitted by node 23 (Alice) as the reference signal vector in the initialization phase of Algorithm 1. The ISE algorithm is based on a non-parametric clustering algorithm and can cluster multiple identifying signatures for classification. The encoder and the decoder of AE have three-layer neural networks with full connection and use the ReLU function as the activation function. The number of hidden variables in AE is 32, and the number of hidden variables in the AE module and the VAE module of TF-HVAE is 64 and 32, respectively.

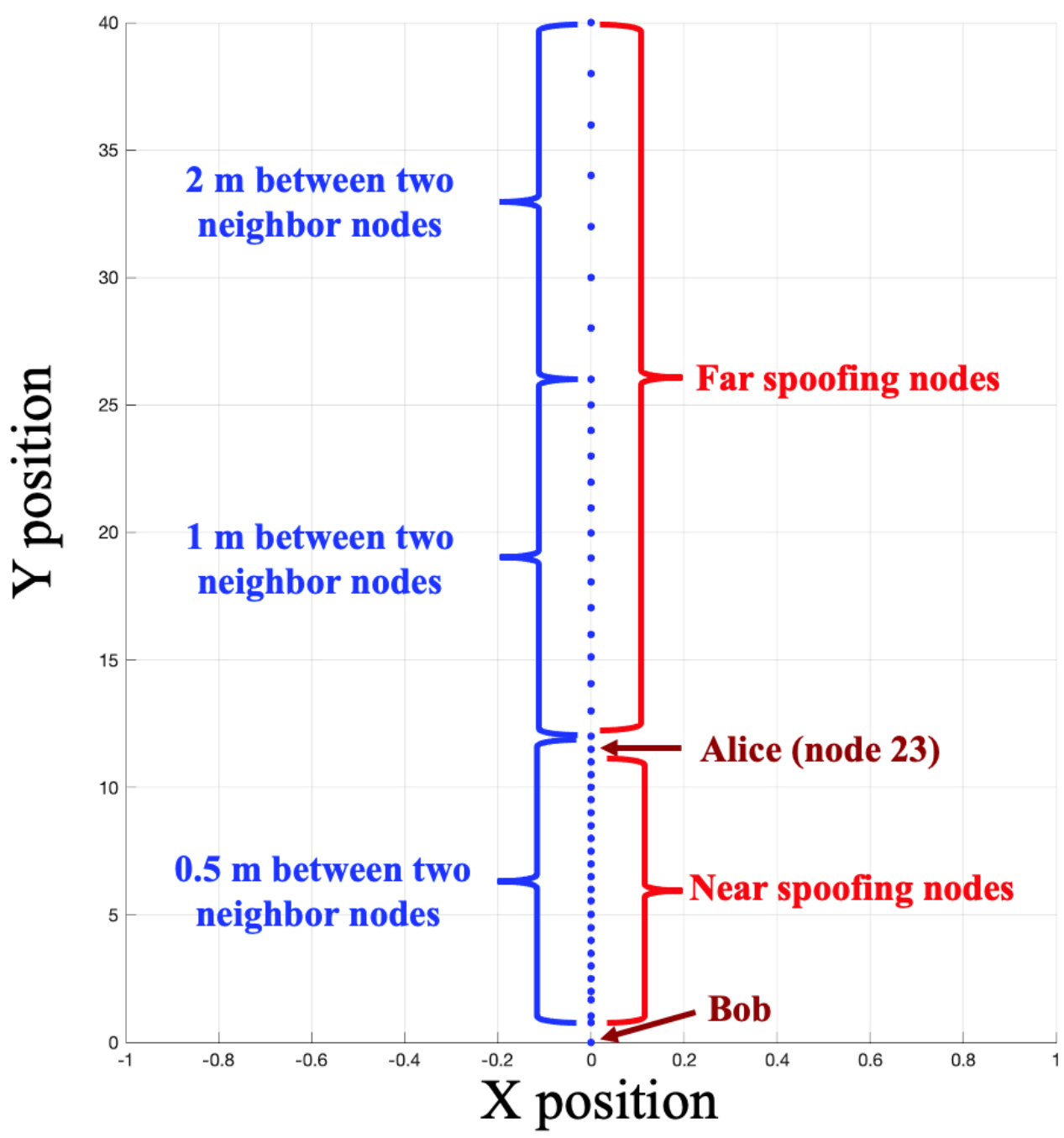

Fig. 3. The position of all nodes in the Open Area Test Site (OATS), where the node 23 and the node 0 are denoted as legal transmitting node (Alice) and Bob, respectively, and the others are represented as spoofing nodes.

TABLE III
THE SIMULATION PARAMETERS IN THE OATS

| Parameters | Values |
|---|---|
| Pilot signals | 2,047 PN Code |
| Transmitting power | 3.64 W |
| Transmitting antenna gain | 10.32 dBi |
| Transmitting antenna | Horn antenna |
| Receiving antenna gain | -3.5 dBi |
| Receiving antenna | Omni-directional |
| Carrier frequency | 5.4 GHz |
| Oversampling number | 4 |
| Sample rate | 200 MHz |
| Expected value of the path loss exponent | 1.9 |
| Expected value of delay | 24.4 ns |
| Expected value of delay spread | 6.1 ns |
| Expected value of K-factor | 12.2 dB |

Fig. 4 presents the loss in the training process with our proposed Threshold-Free Hierarchical Variational Auto-Encoder (TF-HVAE) algorithm versus the training epoch. When the number of epochs is 31, the loss reaches 0.0002, which verifies that the CIR feature extraction by the TF-HVAE can achieve convergence quickly.

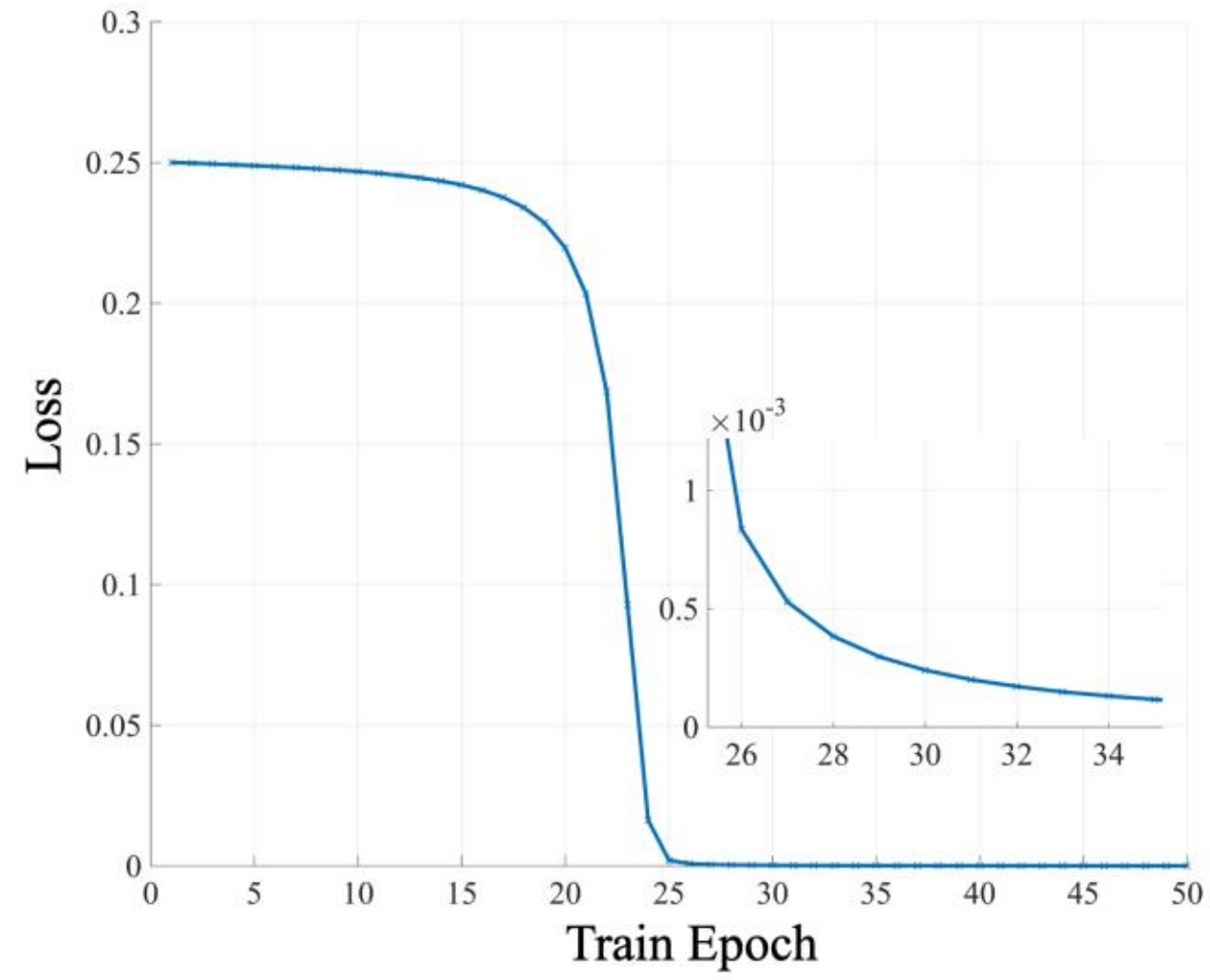

Fig. 4. The loss in the training process with TF-HVAE versus train epoch.

Considering the performance of the PLA model rely on the similarity between the malicious and the normal signal, node 23 is chosen as Alice since it has the closest distance to the nearest neighboring node. In addition, it has the smallest average distance to the remaining nodes, which brings more challenges to the authentication task. Therefore, Fig. 5 demonstrates the lower bound of the proposed scheme. Fig. 5 analyzes the $F_1$ measure of spoofing nodes under different algorithms in the OATS. The computational complexity of ISE increases exponentially with the dimension of CIR [19], and the CIR measured by NIST is the complex vector of 8,188 dimensions, so the classification for CIRs by the clustering algorithm is not suitable for IIoT. [19] used ISE for clustering multiple correlated attributes, which are Channel Gain (CG), channel phase (CP), peak power (PP), and root mean square delay spread (RMS-DS) and can be obtained by CIR. As shown in Fig. 5, the Threshold-Free Auto-Encoder (TF-AE) has much higher authentication performance than the ISE when the spoofing nodes are near Alice, especially node 19 to node 28. The ISE algorithm can only cluster CG, CP, PP, and RMS-DS, representing the coarse-grained information of wireless channels, but can not cluster CIR. However, CIR represents the fine-grained characteristics of wireless channels. Therefore, there is room to improve the authentication performance of the ISE algorithm. In contrast, AE and HVAE can extract CIR features and select valuable features for classification and authentication through neural networks. Threshold-Based Auto-Encoder (TB-AE) has better authentication performance than TF-AE because TB-AE can obtain the best threshold through hypothesis testing. The TF-HVAE has the best authentication performance, which indicates the proposed hierarchical architecture with AE and VAE can extract the intrinsic features of CIR without degrading the generation performance, and the designed double-peak Gaussian distribution can improve the classification performance.

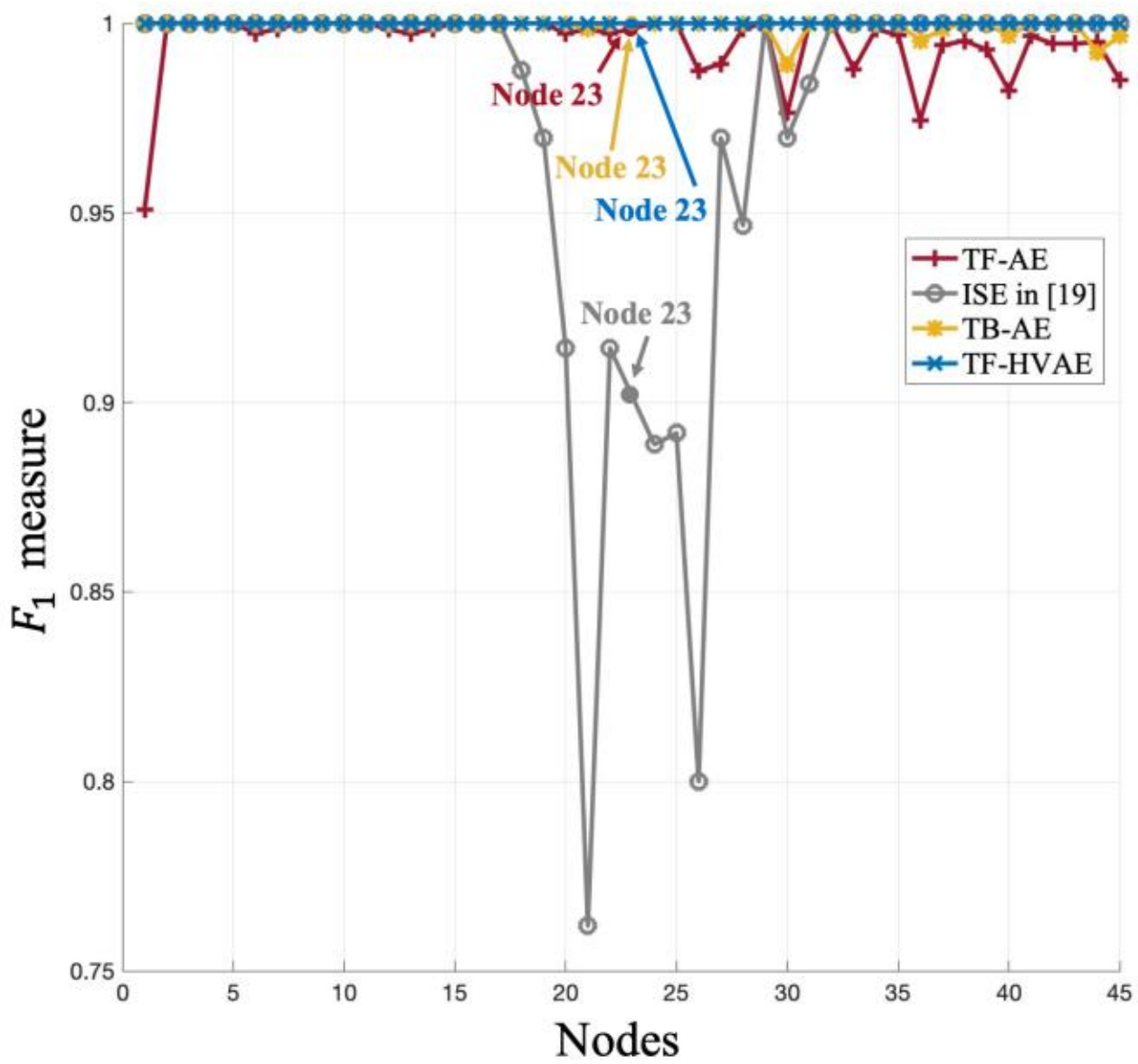

Fig. 5. The authentication performance of spoofing nodes under different algorithms in the OATS, where Alice is located at node 23.

TABLE IV
THE AVERAGE $F_1$ MEASURE IN THE OATS
(ALICE IS LOCATED AT NODE 23)

| Algorithm | Average $F_1$ Measure |
|---|---|
| ISE in [19] | 0.9798 |
| TF-AE | 0.9983 |
| TB-AE | 0.9992 |
| TF-HVAE | 1.0 |

Table IV provides the average $F_1$ measure of spoofing nodes under different algorithms in the OATS. The feature extracting through AE is at least 0.0185 better than ISE [19]. The $F_1$ measure of every spoofing node can reach 1.0 through TF-HVAE, which indicates Bob can authenticate every spoofing node.

Fig. 6 presents the authentication performance of TB-AE in the OATS versus different thresholds. The method of threshold selection refers to [19]. Because the quality of data sets collected in different transmitting nodes is different, TB-AE can choose different thresholds for CIR samples to obtain the best authentication performance. For example, when the 21st-60th CIRs of 400 CIRs per transmitting node are trained and tested by TB-AE, the average $F_1$ measure of 44 spoofing nodes is 1.

Fig. 7 illustrates the authentication performance in the OATS versus different distances between the legitimate node and spoofing nodes. The ISE algorithm has the lowest $F_1$ measure when the spoofing nodes are near Alice. The ISE algorithm makes defending against the spoofing nodes near the legitimate transmitter challenging. The TF-HVAE has better authentication than the other algorithms when the spoofing nodes are near Alice. Our proposed TF-HVAE authentication scheme can defend against the attacks of both near and far spoofing nodes in the static environment.

For more convincing, simulations are carried out when Alice is located on node 1 and 45, respectively. Node 1 is chosen since it has the closest distance to the neighboring devices, making it more vulnerable to neighboring attackers due to relatively more similar CIR. In addition, node 1 is the closest node to Bob, while node 45 is chosen since it is the farthest node to Bob.

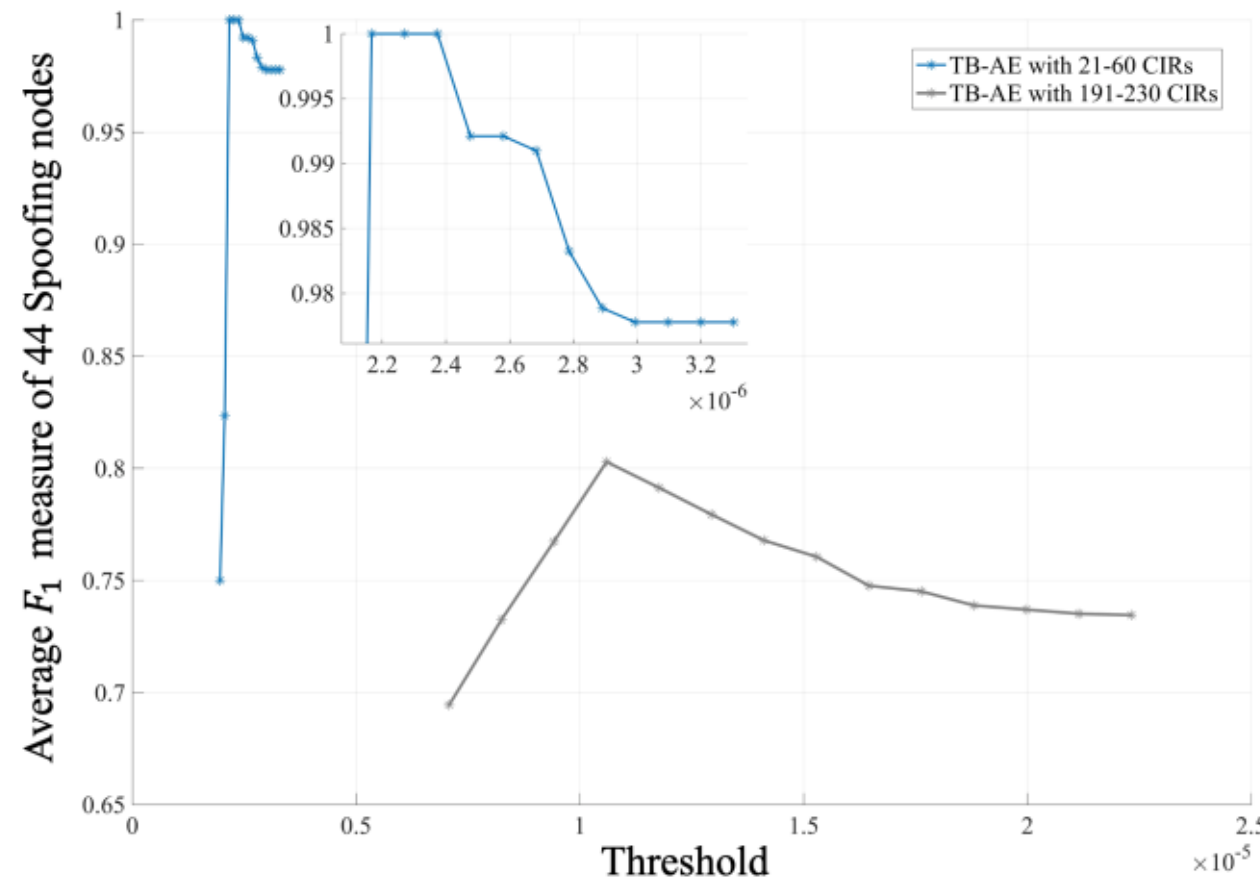

Fig. 6. The authentication performance of TB-AE in the OATS versus different thresholds.

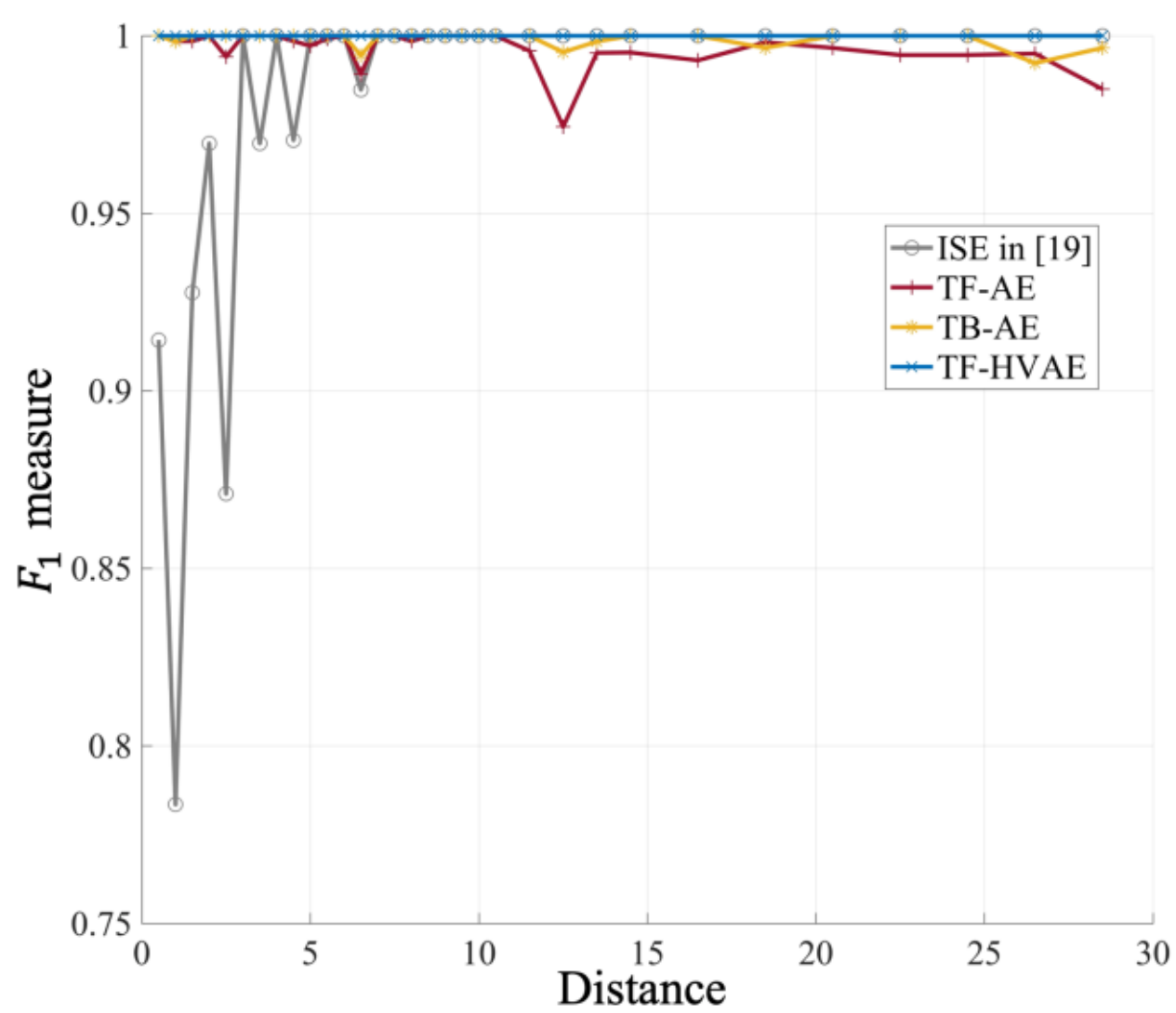

Fig. 7. The authentication performance in the OATS versus different distances between legitimate node and spoofing nodes.

Fig. 8 illustrates the $F_1$ measure of spoofing nodes under different algorithms in the OATS, where Alice is located at the nearest node from Bob, node 1. As the distance between the spoofing node and Alice increases, the authentication performance of TF-AE, ISE, and TB-AE gradually improve. When the spoofing node is close to Alice, node 2 to node 7, the authentication performance of ISE is much worse than the other algorithms. When the spoofing node is far from Alice, node 8 to node 45, the $F_1$ measure of the above algorithms can reach 1. The proposed TF-HVAE algorithm can obtain the best authentication performance, no matter where the attacking node is.

Table V provides the average $F_1$ Measure of spoofing nodes under different algorithms in the OATS, where Alice is located at node 1. Compared with Table V (Alice is located at node 23), fewer spoofing nodes are near Alice. Therefore, ISE, TF-AE, and TB-AE can obtain a higher average $F_1$ Measure. The authentication performance through AE is at least 0.0121 better

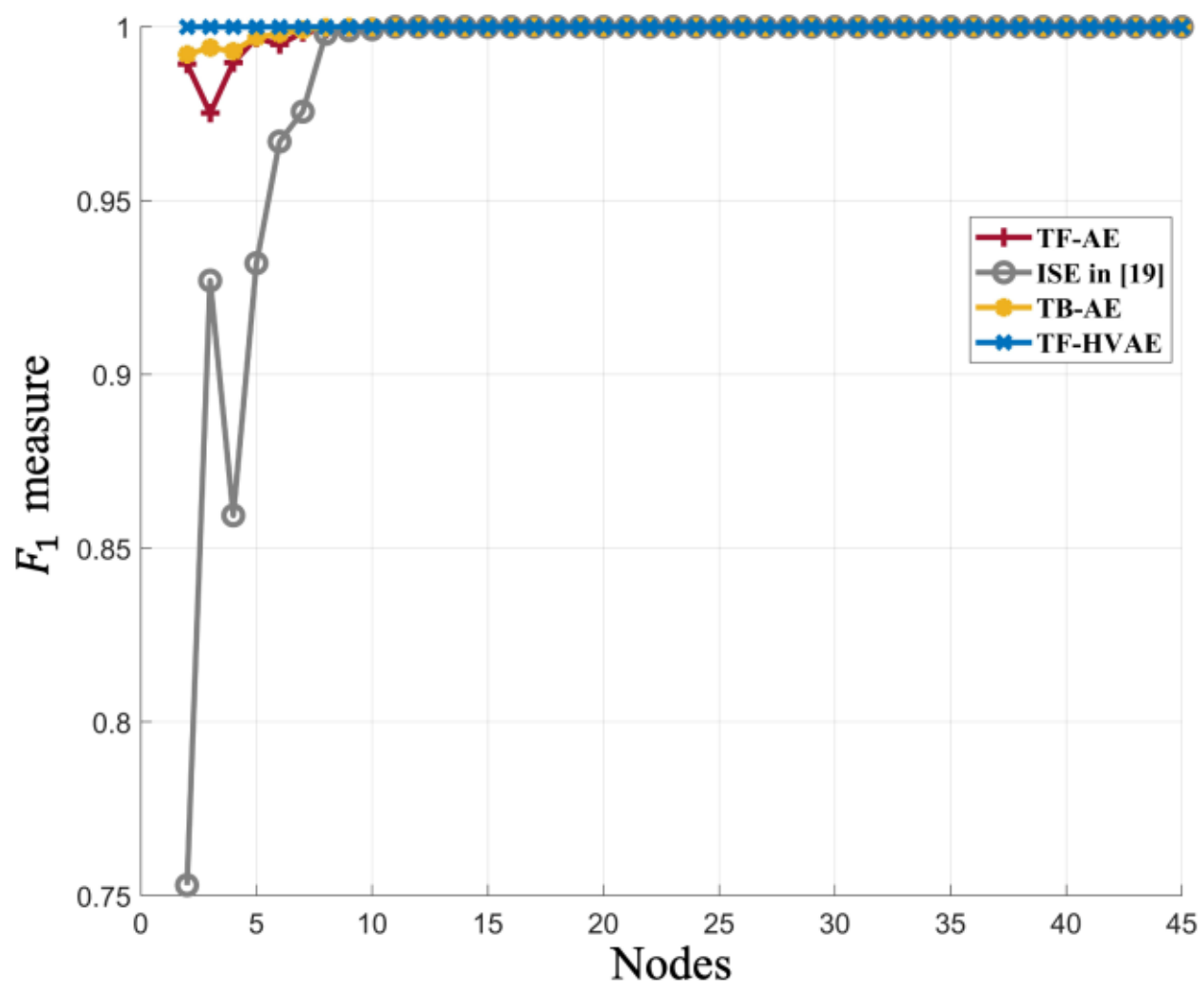

Fig. 8. The authentication performance of spoofing nodes under different algorithms in the OATS, where Alice is located at node 1.

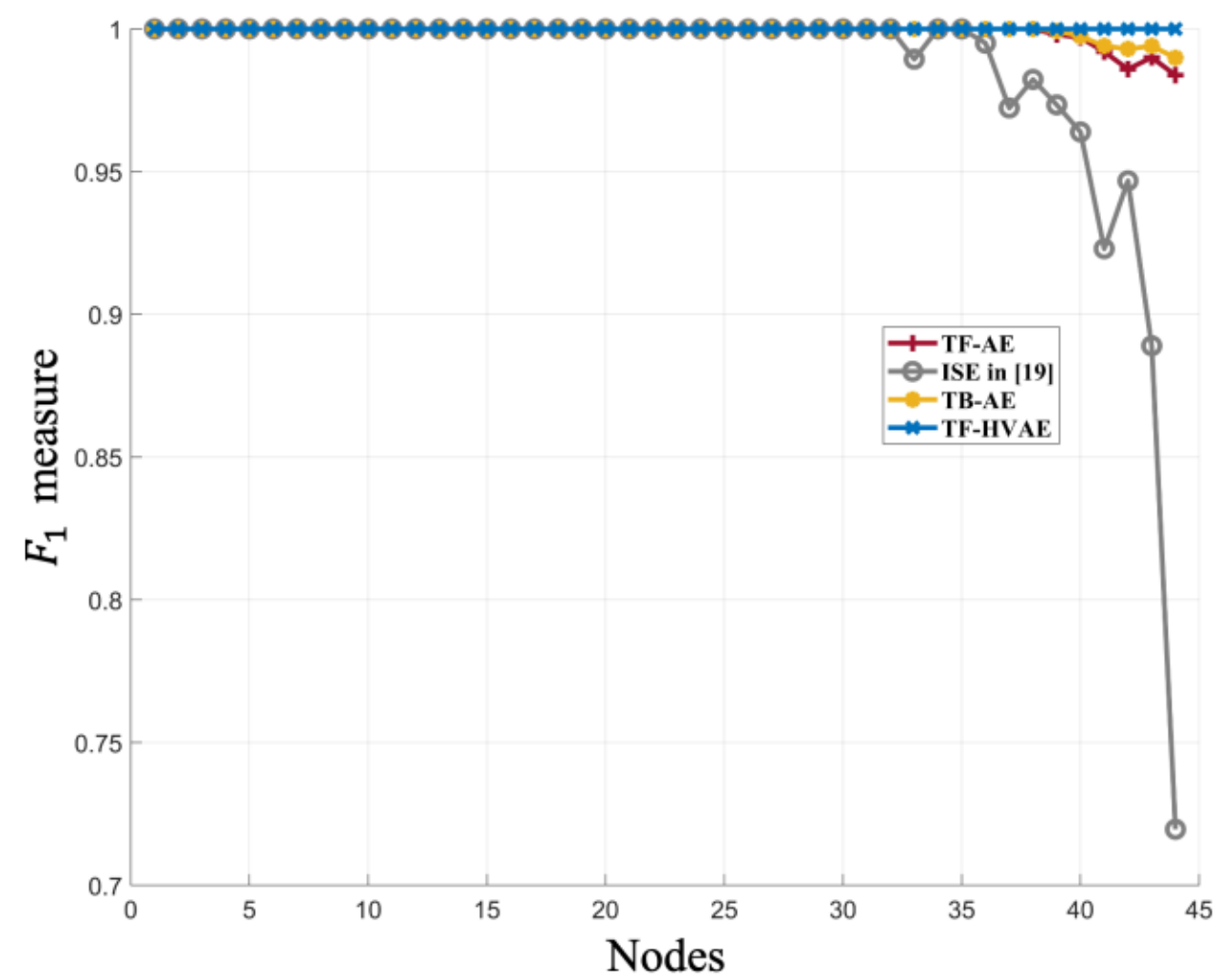

Fig. 9. The authentication performance of spoofing nodes under different algorithms in the OATS, where Alice is located at node 45.

TABLE V
THE AVERAGE $F_1$ MEASURE IN THE OATS
(ALICE IS LOCATED AT NODE 1)

| Algorithm | Average $F_1$ Measure |
|---|---|
| ISE in [19] | 0.9866 |
| TF-AE | 0.9987 |
| TB-AE | 0.9994 |
| TF-HVAE | 1.0 |

TABLE VI
THE AVERAGE $F_1$ MEASURE IN THE OATS
(ALICE IS LOCATED AT NODE 45)

| Algorithm | Average $F_1$ Measure |
|---|---|
| ISE in [19] | 0.9853 |
| TF-AE | 0.9985 |
| TB-AE | 0.9992 |
| TF-HVAE | 1.0 |

than ISE, and Bob can authenticate every spoofing node utilizing TF-HVAE.

Fig. 9 analyzes the $F_1$ Measure of spoofing nodes under different algorithms in the OATS, where Alice is located at the farthest node from Bob, node 45. Similar to the analysis of Fig. 5 and 8, the $F_1$ Measure of TF-AE, ISE, and TB-AE gradually improve as the distance between Alice and the spoofing node increases. When the spoofing node is close to the legitimate node, node 37 to node 44, the authentication performance of ISE is much worse than the other algorithms. When the spoofing node is far from the legal node, node 1 to node 32, the $F_1$ Measure of the above algorithms can reach 1. The proposed TF-HVAE algorithm can authenticate all the spoofing nodes.

Table VI provides the average $F_1$ Measure of spoofing nodes under different algorithms in the OATS, where Alice is located at node 45. From Fig. 5, we can see that the channel differences of all nodes between node 37 to node 44 extracted by AE are smaller than those of node 2 to node 7. Therefore, although the average $F_1$ Measure in Table VII is higher than that in Table V (Alice is located at node 23), it is lower than that in Table VI (Alice is located at node 1). Tables V, VI, and VII verify the superiority of the proposed TF-HVAE in authentication performance.

*2) Simulations with the AF:*

The automotive factory (AF) is a large-scale indoor factory environment. The size of AF is 400 m × 400 m × 12 m, and the AF is a typical Rayleigh fading channel with multiple non-LOS paths. As illustrated in Fig. 10, the location of Bob is (274.17, 151.64), and the transmitter (Alice) moved from (272.8874, 150.876) at an average speed of 1.1 m/s and returned to the starting point along a loop. The loop consists of 121 nodes, some sight lines and some non-sight lines between two nodes, and every node has 300 CIR records. To simulate authentication performance on these data sets, we assumed that the spoofing node moved along the same path as Alice at the same speed, but the start time is a little later. Due to the transmitter's unexpected pause and slow movement during the measurement, we selected 32 nodes as the effective nodes. Table VII shows the simulation parameters in the AF.

In the scenario of the AF, we used 30 CIR samples per node for training and 10 CIR samples per node for testing. There are 32 transmission nodes, so we trained 960 (30 × 32) CIRs with Auto-Encoder (AE), Variational Auto-Encoder (VAE), and Hierarchical Variational Auto-Encoder (HVAE), respectively, and then tested the remaining 10 CIRs of each node to obtain the $F_1$ measure of 17 spoofing nodes. For example, when Alice moves to node 2, the spoofing node moves to node 1. We give the unique label to the first CIR transmitted by node two as the reference signal vector in the initialization phase of Algorithm 1 and obtain the $F_1$ measure of node 1. The encoder and the decoder of both AE and VAE have three-layer neural networks with full connection and use the ReLU function as the activation function. The number of hidden variables in both AE and VAE is 32, and the number of hidden variables in the AE module and the VAE module of TF-HVAE is 64 and 32, respectively.

Due to the long multipath delay and large Doppler shift in a dynamic indoor environment, it is challenging to design a suitable neural network to improve authentication performance. Fig. 11 compares the authentication performance of spoofing nodes under different algorithms in the AF. TF-AE has the worst authentication performance for most spoofing nodes, but TF-VAE has worse performance for some nodes (node 23 to node 29). Section III states that AE encodes CIR samples as points, so the authentication performance is limited in a dynamic environment. Although VAE encodes CIR samples as Gaussian distribution, it can cause over-fitting with few samples, thus reducing the authentication performance. The best threshold for authentication can be obtained by hypothesis testing, which ensures that TB-AE can obtain higher authentication performance than TF-AE. The proposed TF-HVAE first uses the AE module to reduce dimension and then uses the VAE module for further feature extraction, thus extracting the intrinsic characteristics of time-varying channels while avoiding over-fitting. Further, the revised double-peak Gaussian distribution guarantees the authentication performance of unsupervised learning, so the $F_1$ measure of TF-HVAE is always higher than the other algorithms. Table VIII analyzes the average $F_1$ measure of spoofing nodes under different algorithms in the AF. The proposed TF-HVAE can improve the authentication performance by 17.18%, 31.4%, and 69.3% compared with TB-AE, TF-VAE, and TF-AE, respectively. Fig. 11 and Table VIII demonstrate the superiority of the double-peak Gaussian-based PLA explicitly.

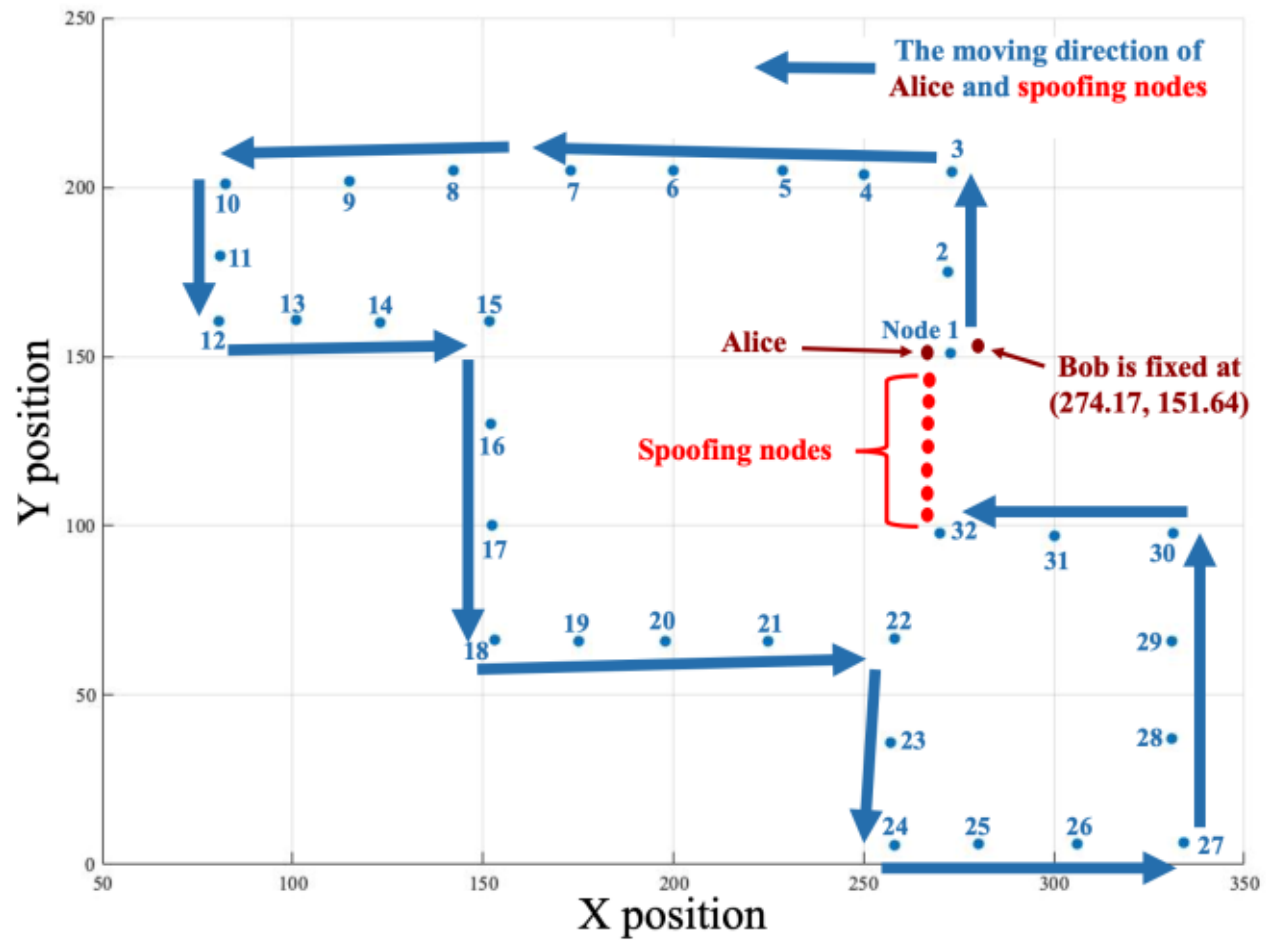

Fig. 10. The position of all nodes in the Automotive Factory (AF), where the spoofing nodes follow Alice's movement.

TABLE VII
THE SIMULATION PARAMETERS IN THE AF

| Parameters | Values |
|---|---|
| Pilot signals | 2,047 PN Code |
| Transmitting power | 1.25 W |
| Transmitting antenna gain | 3.6 dBi |
| Transmitting antenna | Omni-directional |
| Receiving antenna gain | -3.5 dBi |
| Receiving antenna | Omni-directional |
| Carrier frequency | 5.4 GHz |
| Oversampling number | 4 |
| Sample rate | 80 MHz |
| Expected value of the path loss exponent | 3.6 |
| Expected value of delay | 644.4 ns |
| Expected value of delay spread | 177.4 ns |
| Expected value of K-factor | 4.7 dB |

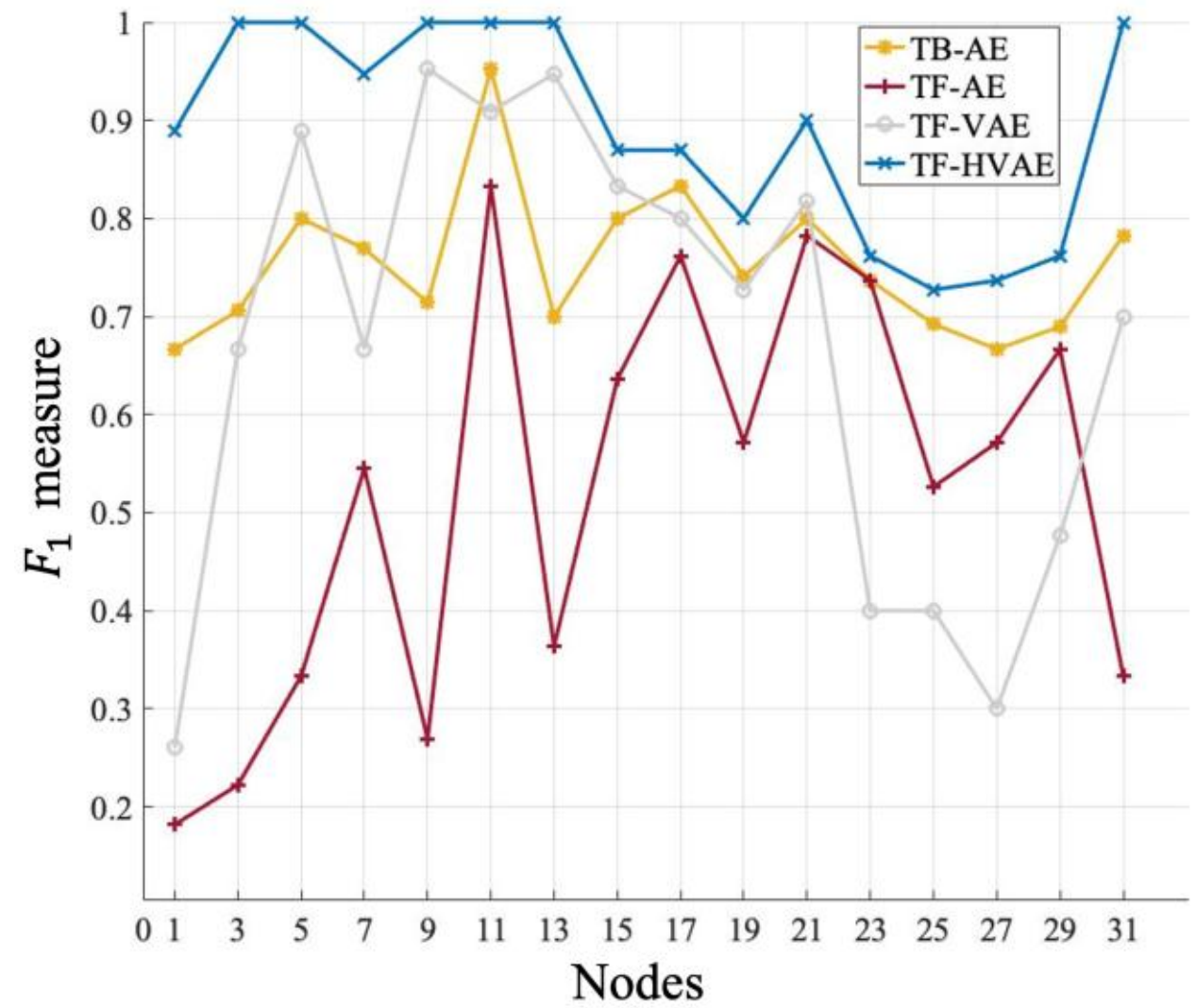

Fig. 11. The authentication performance of spoofing nodes under different algorithms in the AF.

TABLE VIII
THE AVERAGE $F_1$ MEASURE IN THE AF

| Algorithm | Average $F_1$ Measure |
|---|---|
| TF-AE | 0.5210 |
| TB-AE | 0.7531 |
| TF-VAE | 0.6716 |
| TF-HVAE | 0.8825 |

Fig. 12 shows the influence of whether to update the training samples on authentication performance. With the authentication progress, the training sample sets of the online authentication are updated. The training sample sets of the offline authentication are the CIRs collected from Node 1 and Node 2, even when authenticating Node 3 to Node 32. Node 3 is near Node 1, which indicates the data sets of Noe 1 and Node 3 have strong similarities, so the $F_1$ measure of offline authentication is not worse than that of online authentication when authenticating Node 3. With the movement of Alice and the spoofing node, the similarity of sample features between subsequent nodes and Node 1 continues to decline, so the authentication performance of offline authentication for Node 5 to Node 19 is reduced. As the nodes approach the starting point along the loop, the similarity of sample features between authentication nodes and Node 1 increases, so the authentication performance of offline authentication for Node 23 to Node 31 is improved.

Fig. 13 indicates the authentication performance of the TF-HVAE in the AF versus different distances between spoofing nodes and Alice. The interval is defined as the distance interval between the spoofing node and Alice when it starts to move. The interval equals to 2 indicates the spoofing node starts to move when Alice has moved two nodes. For example, when Alice moves to Node 3, the spoofing node moves to Node 1. The average $F_1$ measure is defined as the average authentication performance of spoofing nodes. $h$ and $z$ denote the number of hidden variables in the AE module and the VAE module of TF-HVAE in Fig. 2, respectively. As can be seen, the average $F_1$ measure under $h = 64, z = 32$ reaches 0.96 when the interval is more than 6. The average $F_1$ measure under $h = 64, z = 32$ and $h = 128, z = 64$ is near 1.0 when the interval is more than 9. Fig. 12 and Fig. 13 confirm that the proposed TF-HVAE authentication scheme can obtain good authentication performance in the dynamic indoor environment.

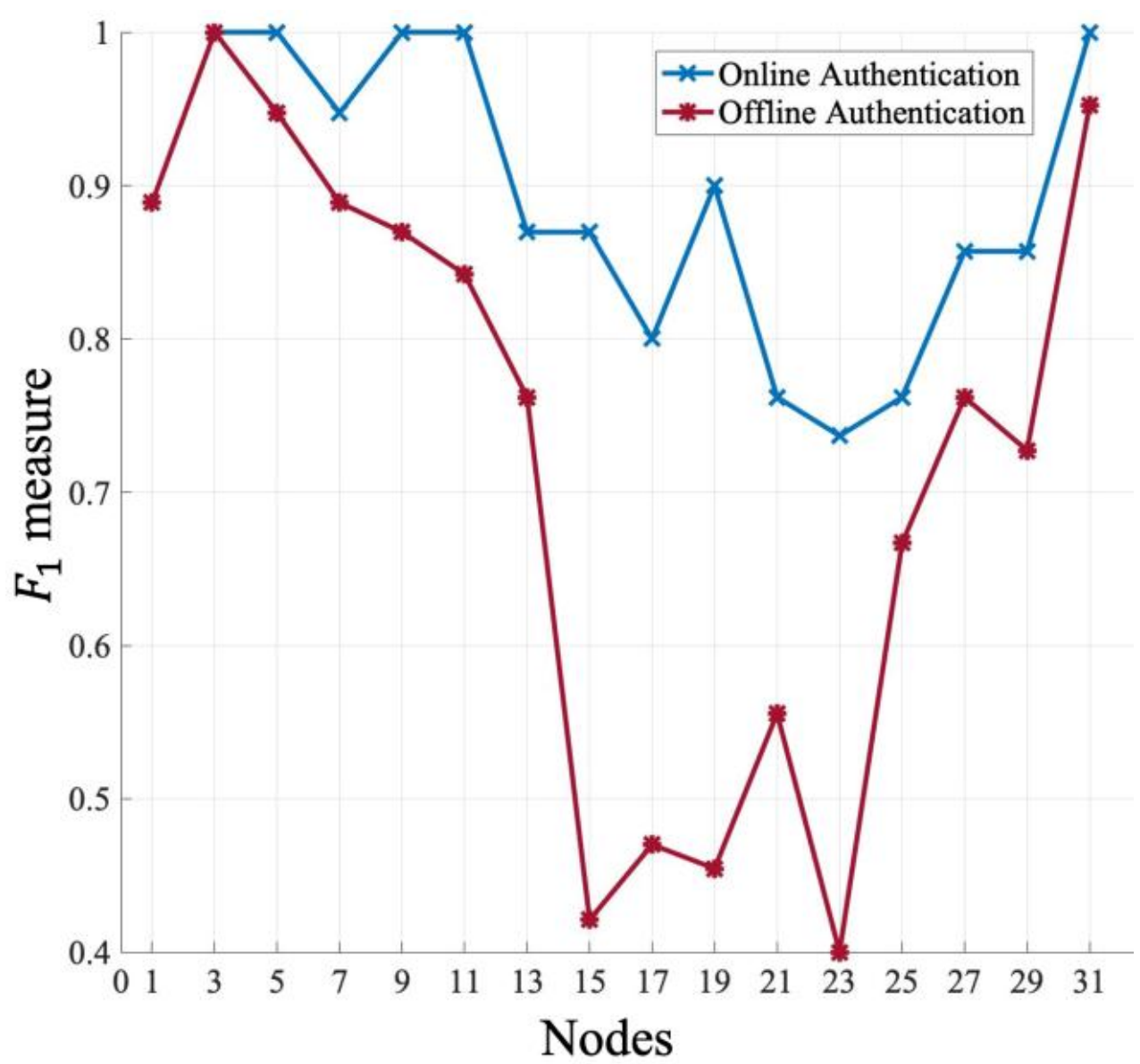

Fig. 12. The online authentication performance and offline authentication under TF-HVAE in the AF.

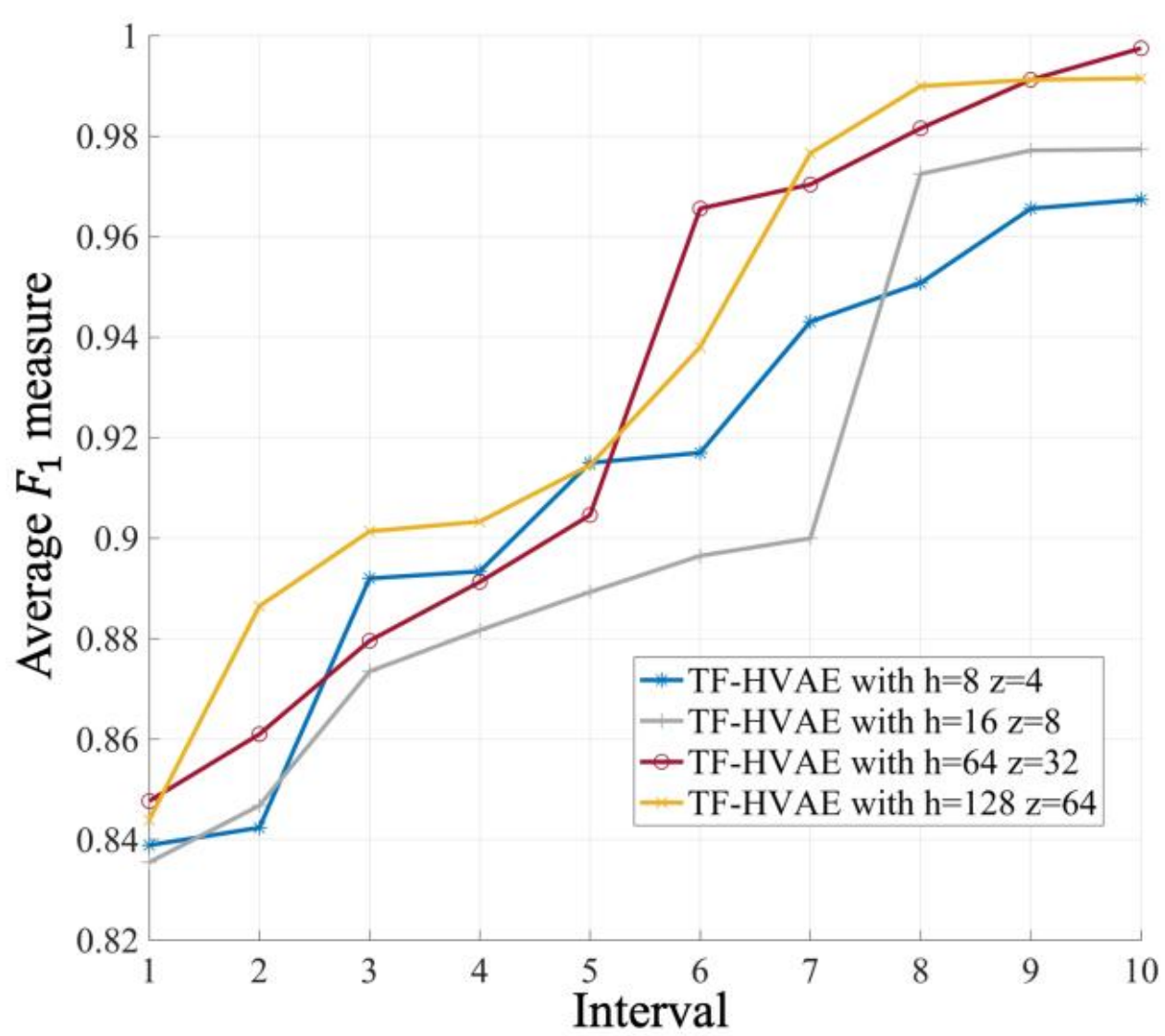

Fig. 13. The authentication performance of the TF-HVAE in the AF versus different distances between spoofing nodes and Alice.

TABLE IX
THE AVERAGE $F_1$ MEASURE IN THE AF

| Parameters of TF-HVAE | Average $F_1$ Measure of different intervals |
|---|---|
| $h = 8, z = 4$ | 0.9126 |
| $h = 16, z = 8$ | 0.9050 |
| $h = 64, z = 32$ | 0.9291 |
| $h = 128, z = 64$ | 0.9340 |

Table IX further illustrates the specific authentication performance under different parameters of TF-HVAE. When AE and VAE in TF-HVAE have fewer hidden variables, the training of neural networks requires more samples. However, the number of wireless signals transmitted in a particular area is limited due to the high speed of authentication nodes in the actual authentication process. As a result, the TF-HVAE with $h = 128, z = 64$ can obtain the best authentication performance in actual PLA.

As illustrated in Table IV and VIII, the average $F_1$ measure of TF-HVAE in the OATS can reach 1.0. While the average $F_1$ measure of TF-HVAE in the AF is 0.8825. Similar to other PLA works [12] [14], the proposed model's performance in the AF scenario is much worse than that in the OATS scenario. Machine-learning-based algorithms' performance relies on the quality of the data and the similarity between the training and the testing data set. The OATS is an open-range facility for calibration and reference measurement of radio wave propagation. It is located in an elevated open space with minimal interference from accidental transmitters and controllable multi-channel components. The AF is an indoor transmission assembly factory covering more than acres. The open area of the factory includes a machining area, inspection machine, assembly workshop, and stacking storage area. Therefore, the performance of the proposed model in the AF scenario is much worse than that in the OATS scenario considering the low quality of data caused by the deep fading and severe multipath effects (reflected by the low K-factor in AF shown by the comparison between Table III and VII).

According to the measurements provided by [38], K-factor is usually defined as the ratio of LOS power to the mean power of diffuse components, which can be used to measure the quality of the channel. Besides, the mobility of the transmitters in the AF scenario leads to a large Doppler Shift and the distribution derivation of the testing and the training data, leading to further performance degradation.

## VI. CONCLUSIONS

In order to defend against spoofing attacks in the IIoT, we propose an HVAE-based PLA scheme to achieve identity authentication. Compared with the cryptography-based authentication schemes, the proposed scheme does not require key interaction and does not rely on the computing resource of transmitters, thus achieving lightweight authentication. To introduce DL into PLA, we prove that the PLA can be approximated with an arbitrarily small error by a neural network. Unlike existing DL-based authentication schemes, the designed scheme does not require attackers' prior channel information, consequently more generality. The constructed loss function considers both the reconstruction error and the representation quality of CIR, thus expanding the depth of identity and improving the authentication performance. $F_1$ Measure is used to metric the authentication performance to obtain more comprehensive authentication results. The simulation results on real industrial data sets indicate that the proposed HVAE algorithm is superior to the unsupervised clustering-based method in authentication for near spoofing nodes in a static outdoor environment. Besides, the HVAE-based scheme can obtain better authentication performance than vanilla AEs & VAEs by 17.18% to 69.3% in a dynamic indoor environment. Therefore, the proposed authentication scheme can be used for identity authentication to ensure wireless security in IIoT. The proposed methods can be theoretically generalized to the multi-user authentication scenario by taking the place of the double-peak Gaussian distribution-based classification module in VAE with multiple-peaks Gaussian distribution-based module. However, we believe that there is a trade-off between the involved class number and the authentication performance, which will be discussed in the future.

## APPENDIX A.
## THE PROOF OF FORMULA (30)

$$
\begin{aligned}
&KL(Q_\vartheta(Z_1|H)||P(Z_1|H)) \\
&= \int Q_\vartheta(Z_1|H)\log\frac{Q_\vartheta(Z_1|H)}{P(Z_1|H)}dZ_1 \\
&= \mathbb{E}_{Z_1\sim Q_\vartheta(Z_1|H)}\left[\log\frac{Q_\vartheta(Z_1|H)}{P(Z_1|H)}\right] \\
&= \mathbb{E}_{Z_1\sim Q_\vartheta(Z_1|H)}[\log Q_\vartheta(Z_1|H) - \log P(Z_1|H)] \\
&= \mathbb{E}_{Z_1\sim Q_\vartheta(Z_1|H)}\left[\log Q_\vartheta(Z_1|H) - \log\frac{P(H|Z_1)P(Z_1)}{P(H)}\right] \\
&= \mathbb{E}_{Z_1\sim Q_\vartheta(Z_1|H)}[\log Q_\vartheta(Z_1|H) - \log P(Z_1) - \log P(H|Z_1)] + \log P(H) \\
&= KL(Q_\vartheta(Z_1|H)||P(Z_1)) - \mathbb{E}_{Z_1\sim Q_\vartheta(Z_1|H)}[\log P(H|Z_1)] + \log P(H)
\end{aligned}
\tag{47}
$$

## APPENDIX B.
## THE PROOF OF FORMULA (31)

From (8) we can obtain

$$
\begin{aligned}
&\log P(H) - KL(Q_\vartheta(Z_1|H)||P(Z_1|H)) \\
&= \mathbb{E}_{Z_1\sim Q_\vartheta(Z_1|H)}[\log P(H|Z_1)] - KL(Q_\vartheta(Z_1|H)||P(Z_1))
\end{aligned}
\tag{48}
$$

where $\log P(H)$ is the log-likelihood function we need to maximize, and we need to minimize $KL(Q_\vartheta(Z_1|H)||P(Z_1|H))$. Based on the information theory, the true posterior distribution $P(H|Z_1)$ of $Z_1$ is the best encoder of $H$, which contains the most information about $H$, and the performance of reconstructing $H$ with $P(H|Z_1)$ is optimal. Therefore, if $Z_1$ obtained by $Q_\vartheta(Z_1|H)$ can reconstruct $H$ well, it confirms that $Q_\vartheta(Z_1|H)$ is relatively close to $P(H|Z_1)$.

Moreover, we can obtain

$$
\mathbb{E}_{Z_1\sim Q_\vartheta(Z_1|H)}[\log P(H|Z_1)] - KL(Q_\vartheta(Z_1|H)||P(Z_1)) \leq \log P(H)
\tag{49}
$$

To generate samples $\dot{\mathcal{X}}$ more similar to $H$, we add the constraint as

$$
KL(Q_\vartheta(Z_1|H)||\mathcal{N}(0,I)) = KL(P(Z_1|H)||P(Z_1))
\tag{50}
$$

Then we can obtain $P(Z_1)\sim\mathcal{N}(0,I)$ as

$$
\begin{aligned}
P(Z_1) &= \int P(Z_1|H)P(H)\,dH \\
&= \int \mathcal{N}(0,I)P(H)\,dH \\
&= \mathcal{N}(0,I)\int P(H)\,dH \\
&= \mathcal{N}(0,I)
\end{aligned}
\tag{51}
$$

We calculate $KL(Q_\vartheta(Z_1|H)||\mathcal{N}(0,I))$ as

$$
\begin{aligned}
&KL(Q_\vartheta(Z_1|H)||\mathcal{N}(0,I)) \\
&= \int \frac{1}{\sqrt{2\pi\sigma^2}}e^{-\frac{(H-\mu)^2}{2\sigma^2}}\log\frac{\frac{1}{\sqrt{2\pi\sigma^2}}e^{-\frac{(H-\mu)^2}{2\sigma^2}}}{\frac{1}{\sqrt{2\pi\sigma^2}}e^{-\frac{H^2}{2}}}dH \\
&= \int \frac{1}{\sqrt{2\pi\sigma^2}}e^{-\frac{(H-\mu)^2}{2\sigma^2}}\log\frac{1}{\sqrt{\sigma^2}}\times e^{-\frac{H^2}{2}-\frac{(H-\mu)^2}{2\sigma^2}}dH \\
&= \int \frac{1}{\sqrt{2\pi\sigma^2}}e^{-\frac{(H-\mu)^2}{2\sigma^2}}\left[-\frac{1}{2}\log\sigma^2 + \frac{1}{2}H^2 - \frac{1}{2}\frac{(H-\mu)^2}{\sigma^2}\right]dH \\
&= \frac{1}{2}\int \frac{1}{\sqrt{2\pi\sigma^2}}e^{-\frac{(H-\mu)^2}{2\sigma^2}}\left[-\log\sigma^2 + H^2 - \frac{(H-\mu)^2}{\sigma^2}\right]dH \\
&= \frac{1}{2}\left(-\log\sigma^2 + \mathbb{E}[H^2] - \frac{1}{\sigma^2}\mathbb{E}[(H-\mu)^2]\right) \\
&= \frac{1}{2}(-\log\sigma^2 + \sigma^2 + \mu^2 - 1)
\end{aligned}
\tag{52}
$$

## APPENDIX C.
## THE PROOF OF FORMULA (25)

$$
\begin{aligned}
&KL(N(\mu,\sigma^2)||\frac{1}{2}N(-m,s^2) + \frac{1}{2}N(m,s^2) \\
&= KL(2*\frac{1}{2}N(\mu,\sigma^2)||\frac{1}{2}N(-m,s^2) + \frac{1}{2}N(m,s^2)) \\
&\leq KL(\frac{1}{2}N(\mu,\sigma^2)||\frac{1}{2}N(-m,s^2)) + KL(\frac{1}{2}N(\mu,\sigma^2)|| + \frac{1}{2}N(m,s^2)) \\
&= -\frac{1}{2}\log 2\left(s^2 + \log\sigma^2 - \frac{1}{2}(\mu-m)^2 - \frac{1}{2}(\mu+m)^2 - \sigma^2\right)
\end{aligned}
\tag{53}
$$

## APPENDIX D.
## THE PROOF OF FORMULA (26)

$$
KL(N(\mu,\sigma^2)||\frac{1}{2}N(-m,s^2) + \frac{1}{2}N(m,s^2))
$$

$$= \int_{-\infty}^{\infty} N(\mu, \sigma^2) \log \frac{N(\mu, \sigma^2)}{\frac{1}{2} N(-m, s^2) + \frac{1}{2} N(m, s^2)} dx$$

$$= \int_{-\infty}^{\infty} \frac{1}{\sqrt{2\pi\sigma^2}} e^{-\frac{(x-\mu)^2}{2\sigma^2}} \log \frac{\frac{1}{\sigma} e^{-\frac{(x-\mu)^2}{2\sigma^2}}}{\frac{1}{2s}\left[e^{-\frac{(x-m)^2}{2s^2}} + e^{-\frac{(x+m)^2}{2s^2}}\right]} dx$$

$$= \int_{-\infty}^{\infty} \frac{1}{\sqrt{2\pi\sigma^2}} e^{-\frac{(x-\mu)^2}{2\sigma^2}} \log \frac{2s}{\sigma} dx$$

$$- \int_{-\infty}^{\infty} \frac{1}{\sqrt{2\pi\sigma^2}} e^{-\frac{(x-\mu)^2}{2\sigma^2}} \left[\frac{(x-\mu)^2}{2\sigma^2} - \frac{(x-m)^2}{2s^2}\right] dx$$

$$- \int_{-\infty}^{\infty} \frac{1}{\sqrt{2\pi\sigma^2}} e^{-\frac{(x-\mu)^2}{2\sigma^2}} \log\left[1 + e^{-\frac{2mx}{s^2}}\right] dx$$

$$= \iota - \kappa - \gamma \quad (54)$$

where $\iota$, $\kappa$, and $\gamma$ are denoted as

$$\iota = \log \frac{2s}{\sigma} \quad (55)$$

$$\kappa = -\frac{(m-\mu)^2 + \sigma^2 - s^2}{2s^2} \quad (56)$$

$$\gamma \approx -\frac{2m\left[-\sigma e^{-\frac{\mu^2}{2\sigma^2}} + \sqrt{\frac{\pi}{2}\mu Erfc\left(\frac{\mu}{\sqrt{2}\sigma}\right)}\right]}{s^2} \quad (57)$$

With an approximation of $Erfc(x) \approx 1 - tanh(1.19x)$, we can obtain (26).